\documentclass[aps,floats,twocolumn,psfig,epsf,prb,showpacs]{revtex4}
\usepackage{amssymb}
\usepackage{amsmath}
\usepackage{graphicx}

\input{tcilatex}

\begin{document}

\title{The effect of vertex corrections on the possibility of chiral
symmetry breaking, induced by long-range Coulomb repulsion in graphene}
\author{A. Katanin}
\affiliation{Institute of Metal Physics, 620990, Kovalevskaya str. 18 Ekaterinburg, Russia%
\\
Ural Federal University, 620002, Mira str. 19, Ekaterinburg, Russia}

\begin{abstract}
In \ this paper we consider the possibility of chiral (charge or spin
density wave) symmetry breaking in graphene due to long-range Coulomb
interaction by comparing the results of the Bethe-Salpeter and functional
renormalization-group approaches. The former approach performs summation of
ladder diagrams in the particle-hole channel, and reproduces the results of
the Schwinger-Dyson approach for the critical interaction strength of the
quantum phase transition. The renormalization-group approach combines the
effect of different channels and allows to study the role of vertex
corrections. The critical interaction strength, which is necessary to induce
the symmetry breaking in the latter approach is found in the static
approximation to be $\alpha _{c}=e^{2}/(\epsilon v_{F})\approx 1.05$ without
considering the Fermi velocity renormalization, and $\alpha _{c}=3.7$ with
accounting the renormailzation of the Fermi velocity. The latter value is
expected to be however reduced, when the dynamic screening effects are taken
into account, yielding the critical interaction, which is comparable to the
one in freely suspended graphene. We show that the vertex corrections are
crucially important to obtain the mentioned values of critical interactions.
\end{abstract}

\maketitle

\section{Introduction}

The fascinating physics of Dirac fermions, interacting via static Coulomb
potential, became a subject of intensive research in condensed-matter
physics after the experimental realization of graphene \cite{Novoselov}.
Despite vanishing density of states at the Fermi level in clean graphene,
Coulomb interaction may lead to different instabilities of electronic system 
\cite{Herbut,Honerkamp,Rev,Herbut1}. Even before the experimental discovery
of graphene, the possibility of excitonic instability (related to the charge
density wave formation) in this material was predicted theoretically\cite%
{Khveschenko,Gusynin0}. It was argued, that sufficiently strong Coulomb
interaction is able to open a gap in the excitation spectrum analogously to
the chiral symmetry breaking in quantum electrodynamics\cite{CSB}.

The analysis within the Schwinger-Dyson mean-field approach \cite%
{Khveschenko1,Murthy,Gusynin} in the static approximation found critical
Coulomb interaction, which is necessary to open the charge gap, to be $%
\alpha _{c}=e^{2}/(\epsilon v_{F})=1.62$ (see, e.g., discussion in Ref. %
\onlinecite{Gusynin}), where $e$ is the critical charge, $\epsilon $
corresponds to substrate dielectric constant, and $v_{F}$ is the Fermi
velocity. This result is not very far from the result of quantum Monte-Carlo
approach \cite{MC} $\alpha _{c}=1.08$. Inclusion of the dynamic screening of
Coulomb interaction in Refs. \onlinecite{Khveschenko-vel-dyn,Gusynin1}
allowed to improve further the agreement with the numerical results. These
results were however altered recently by considering the effect of
renormalization of Fermi velocity, yielding critical Coulomb interactions,
which depend on the approximation used, namely the critical couplings $%
\alpha _{c}=1.75$ (Ref. \onlinecite{Gonzalez}), $\alpha _{c}=7.65$ (Ref. %
\onlinecite{Popovici}), and $\alpha _{c}=3.1$ (Ref. \onlinecite{Gonzalez1})
were obtained. The latter result seems to be most reliable within the
Schwinger-Dyson approach, since it accounts for the renormalization of
quasiparticle weight and the polarization bubble. These later results are
all larger than the coupling constant of free-suspended graphene, in
agreement with the experimental fact, that freely suspended graphene shows
no chiral instability\cite{Exp}. On the other hand, the important effect of
the other bands for increasing critical Coulomb interaction was emphasized
in Ref. \cite{Katsnelson}.

In the present paper we concentrate on studying the effect of the long-range
part of Coulomb interaction and address the question, in which parameter
range the Schwinger-Dyson approach is applicable. Indeed, this approach,
which can be reformulated as a ladder approximation for electron interaction
vertices, neglects contribution of the channels, other than the
particle-hole one. From previous studies of the two-dimensional Hubbard
model with short-range (on-site) interaction\cite%
{HonerkampRice,Metzner,Katanin,1PIRev,Honerkamp}, it is known that the
contribution of the particle-hole channel is to a large extent compensated
by the particle-particle channel. For the systems with Dirac electronic
dispersion this cancellation is almost perfect, since the electronic Green
function is odd with respect to momentum and frequency. At a first glance,
this can strongly weaken the possibility of excitonic instability in
graphene.

Apart from that, the discussed approaches neglect electron-electric field
vertex corrections, which also appear beyond the ladder approximation, used
in these studies. Although these corrections are expected to have no
infrared divergencies \cite{Kotov,Sodemann}, in the lowest (second) order in
the Coulomb interaction they increase renormalized dielectric constant of
the substrate. Therefore, one could expect that they yield further increase
of the critical value of the Coulomb interaction for the chiral quantum
phase transition.

The possibility of chiral phase transition in graphene was considered
previously within the field-theoretical renormalization group approach\cite%
{Son,Son1,FostAleiner,Herbut2,Sarma}. It was shown that in the one-loop
approximation \cite{Son,Son1,FostAleiner,Herbut2} Coulomb interaction is
irrelevant due to increase of the Fermi velocity with respect to its bare
value. Apart from that, small point-like interactions, which can be added to
test the (in)stability towards chiral symmetry breaking, are also irrelevant
because of the abovementioned cancellation between particle-particle and
particle-hole channels. Therefore, this approach does not yield divergence
of some coupling constants at sufficiently large Coulomb interaction, and
therefore does not yield chiral phase transition. In the two-loop
approximation \cite{Sarma} there is an instability of the field-theoretical
RG flow at $\alpha _{c}=0.78,$ which can be in principle related to a
possibility of chiral phase transition. However, such a scenario is anyhow
different from the two-particle instability considered within the
Schwinger-Dyson approach.

In the study paper we apply the functional renormalization-group (fRG)
approach\cite{HonerkampRice,Metzner,Katanin,1PIRev,Honerkamp} and argue that
considering momentum dependences of the vertices and summing all
contributions from different channels in the one-loop approximation, the so
far neglected effect of the vertex corrections yields at the same time
enhancing the tendency towards excitonic instability, and to a large extent
compensates the above mentioned factors, which weaken the instability. Taken
together, they yield even smaller result for the critical interaction
strength, than in the ladder approach. In the static approximation without
the renormalization of Fermi velocity we obtain $\alpha _{c}=1.05,$ which is
smaller than previous estimates at the same level of approximation within
the Schwinger-Dyson approach. The critical interaction increases to $\alpha
_{c}^{v}=3.7$ when the Fermi velocity renormalization is taken into account.
The latter value of the critical interaction is however expected to be
reduced when the dynamic effects are taken into account, yielding the
critical coupling, which is comparable to the one in the freely suspended
graphene.

The outline of the paper is the following. In Section II we formulate the
model, in Sect. III we discuss reformulation of Dyson-Schwinger approach as
summation of ladder diagrams and extend previously obtained results to the
symmetric phase. In Sect. IV we consider functional renormalization-group
approach, discuss parametrization of the vertices used in the present study,
and present main results for the electron interaction vertices. The
Conclusion is presented in Sect. V.

\section{The model}

To study electron interaction in graphene we consider the model (see, e.g.,
Refs. \cite{Son,Son1,FostAleiner}) 
\begin{eqnarray}
L_{0} &=&-\int d^{3}X\overline{\psi }_{s}(\gamma _{\mu }\partial _{\mu
}+i\gamma _{0}A_{0})\psi _{s}  \notag \\
&&+\frac{1}{4\pi e^{2}}\int d^{3}Xdz\left[ (\partial
_{a}A_{0})^{2}+(\partial _{z}A_{0})^{2}\right]  \label{L0X}
\end{eqnarray}%
where $\psi _{s}\equiv \psi _{s}(X)$ is the 4-spinor describing electronic
degrees of freedom, $s=\uparrow ,\downarrow $ is the spin projection, $%
X=(\tau ,x,y),$ $\mu =0,1,2,$ $a=1,2,$ $\partial =(\partial _{\tau
},v_{F}\partial _{x},v_{F}\partial _{y}),$ $\overline{\psi }=\psi ^{+}\gamma
_{0},\ \gamma _{\mu }$ are the Dirac matrices, e.g.%
\begin{equation}
\gamma _{0}=\left( 
\begin{array}{cc}
\sigma _{3} & 0 \\ 
0 & -\sigma _{3}%
\end{array}%
\right) ,\ \ \gamma _{a}=\left( 
\begin{array}{cc}
\sigma _{a} & 0 \\ 
0 & -\sigma _{a}%
\end{array}%
\right) ,
\end{equation}%
where $\sigma _{i}$ are the Pauli matrices, $\alpha =e^{2}/(\epsilon v_{F})$
is the dimensionless coupling constant. The quadratic part of the action (%
\ref{L0X}) represents the continuum limit of the microscopic tight-binding
model of graphene (see, e.g., Ref. \cite{FostAleiner}) and corresponds to
the representation $\psi =\{\psi _{A}^{1},\psi _{B}^{1},\psi _{B}^{2},-\psi
_{A}^{2}\}$ ($\psi _{s}^{m}$ is an annihilation operator for the electron in
valley $m$ and sublattice $s$). The scalar field $A_{0}$ accounts for the
Coulomb interaction of Dirac particles, which is integrated along the third
coordinate $z$, transverse to the graphene layer. Excluding $z$ direction,
perpendicular to the graphene layer, and performing Fourier transformation,
yields the effective action\cite{Son,Son1,FostAleiner} 
\begin{eqnarray}
L_{0} &=&-\int \frac{d^{3}pd^{3}q}{(2\pi )^{6}}\overline{\psi }_{ps}\left[
i\gamma _{\mu }p_{\mu }\delta (q)+i\gamma _{0}A_{0q}\right] \psi _{p+q,s} 
\notag \\
&&+\frac{1}{4\pi \alpha }\int \frac{d^{3}q}{(2\pi )^{3}}|\mathbf{q}%
|A_{0q}A_{0,-q}  \label{L0}
\end{eqnarray}%
where $p_{\mu }=(\nu /v_{F},p_{x},p_{y})$ is the $3$-vector, including
Matsubara frequency $\nu $ (and similar for $q$). It is well known (see,
e.g., Ref. \cite{Rev}), that due to vanishing density of states, the Coulomb
interaction in graphene is not screened, since the particle-hole bubble 
\begin{equation}
\Pi (\mathbf{q},\omega )=\frac{q^{2}}{4\sqrt{q^{2}+(\omega /v_{F})^{2}}}
\end{equation}%
(where $q=|\mathbf{q}|$) is itself linear in momentum at $\omega =0.$
Therefore, summation of the bubble contributions in a rundom phase
approximation (RPA) yields renormalization of the dielectric constant only,
such that the effective "screened\textquotedblright\ Coulomb interaction is 
\begin{equation}
\alpha _{r}=\alpha /(1+\pi \alpha /2).  \label{ar}
\end{equation}

\section{The ladder (Bethe-Salpeter) approach without Fermi velocity
renormalization}

The Coulomb interaction may lead to excitonic (charge or spin density wave)
instability with opening the gap in the electronic spectrum \cite%
{Khveschenko,Gusynin0,Khveschenko1,Murthy,Gusynin}. The simplest way to
treat this possibility is the Bethe-Salpeter approach. Here we consider the
instability from the symmetric phase, which complements previous studies of
symmetry-broken phase within the Schwinger-Dyson approach\cite%
{Khveschenko,Gusynin0,Khveschenko1,Murthy,Gusynin}. The (linearized) gap
equation of the Schwinger-Dyson approach in the static approximation (\ref%
{ar}) has the form 
\begin{equation}
\Delta _{\mathbf{p}}=\Delta _{0}+2\pi \alpha _{r}\int\limits_{0}^{\Lambda }%
\frac{d^{2}\mathbf{k}}{(2\pi )^{2}}\mathrm{Tr}\left[ \frac{\gamma
_{0}\otimes \gamma _{0}}{|\mathbf{p}-\mathbf{k}|}\widetilde{\Pi }_{\mathrm{ph%
}}(\mathbf{k},\mathbf{0})\right] \Delta _{\mathbf{k}},  \label{Dp}
\end{equation}%
where we have introduced the bubble 
\begin{eqnarray}
\widetilde{\Pi }_{\mathrm{ph,pp}}(\mathbf{k},\mathbf{q}) &=&-\int \frac{%
d\varepsilon }{2\pi }\left[ G(\mathbf{k},i\varepsilon )\otimes G(\pm \mathbf{%
k+q},\pm i\varepsilon )\right]  \notag \\
&=&\pm \frac{v_{F}}{2(v_{F}(|\mathbf{k}|)|\mathbf{k}|+v_{F}(|\mathbf{k\pm q}%
|)|\mathbf{k\pm q}|)}  \notag \\
&\times &\left[ \gamma _{0}\otimes \gamma _{0}+\frac{k_{a}(k_{b}\pm q_{b})}{|%
\mathbf{k}||\mathbf{k\pm q}|}\gamma _{a}\otimes \gamma _{b}\right]
\label{Pi}
\end{eqnarray}%
(the bubble $\widetilde{\Pi }_{\mathrm{pp}}$ will be used later, in Sect.
IV), $G(\mathbf{k},\varepsilon )=\left( \gamma _{0}\varepsilon +\mathrm{i}%
v_{F}(|\mathbf{k|})\gamma _{a}k_{a}\right) ^{-1}$, the matrices in the
direct product act in the space of spinor indices of each fermionic line, $%
\Delta _{0}$ is the bare value of the gap, and in general the renormalized
Fermi velocity $v_{F}(|\mathbf{k}|)$ depends on momentum. In the present
Section we neglect the renormalization of the Fermi velocity by taking $%
v_{F}(|\mathbf{k|})=v_{F}$. The Eq. (\ref{Dp}) can be iterated to express
the resulting gap 
\begin{equation}
\Delta _{\mathbf{p}}=\Delta _{0}\left\{ 1+\int\limits_{0}^{\Lambda }\frac{%
d^{2}\mathbf{k}}{(2\pi )^{2}}\mathrm{Tr}\left[ \Gamma _{\mathbf{pk0}}%
\widetilde{\Pi }_{\mathrm{ph}}(\mathbf{k},\mathbf{0})\right] \right\}
\label{Dp1}
\end{equation}%
through the vertex function $\Gamma _{\mathbf{kk}^{\prime }\mathbf{q}},$ for
which we obtain the integral equation 
\begin{eqnarray}
\Gamma _{\mathbf{kk}^{\prime }\mathbf{q}} &=&\frac{2\pi \alpha _{r}}{|%
\mathbf{k}-\mathbf{k}^{\prime }|}\gamma _{0}\otimes \gamma _{0} \\
&+&\gamma _{0}\otimes \gamma _{0}\int \frac{d^{2}p}{(2\pi )^{2}}\frac{2\pi
\alpha _{r}}{|\mathbf{k}-\mathbf{p}|}\widetilde{\Pi }_{\mathrm{ph}}(\mathbf{p%
},\mathbf{q})\Gamma _{\mathbf{pk}^{\prime }\mathbf{q}},  \notag
\end{eqnarray}%
corresponding to the summation of ladder diagrams in the particle-hole
channel. At $\mathbf{q}=0$ we therefore have%
\begin{eqnarray}
\Gamma _{\mathbf{kk}^{\prime }\mathbf{0}} &=&\frac{2\pi \alpha _{r}}{|%
\mathbf{k}-\mathbf{k}^{\prime }|}\gamma _{0}\otimes \gamma _{0}+\frac{\alpha
_{r}}{4}\int dp\int \frac{d\theta _{p}}{2\pi }\frac{1}{|\mathbf{k}-\mathbf{p}%
|}  \notag \\
&\times &\left( I\otimes I+\frac{p_{a}p_{b}}{|\mathbf{p}|^{2}}\gamma
_{0}\gamma _{a}\otimes \gamma _{0}\gamma _{b}\right) \Gamma _{\mathbf{pk}%
^{\prime }\mathbf{0}}.  \label{EqGamma}
\end{eqnarray}

We search the solution to this equation in the form%
\begin{eqnarray}
\Gamma _{\mathbf{kk}^{\prime }\mathbf{0}} &=&\Gamma _{\mathbf{kk}^{\prime }%
\mathbf{0}}^{0}\gamma _{0}\otimes \gamma _{0}+\Gamma _{\mathbf{kk}^{\prime }%
\mathbf{0}}^{1}\gamma _{a}\otimes \gamma _{a}  \notag \\
&+&\Gamma _{\mathbf{kk}^{\prime }\mathbf{0}}^{12}i\gamma _{0}\gamma
_{1}\gamma _{2}\otimes i\gamma _{0}\gamma _{1}\gamma _{2}.  \label{Anz}
\end{eqnarray}%
Performing matrix multiplication (see Appendix A), for the s-component of
the vertices $\Gamma _{\mathbf{kk}^{\prime }\mathbf{0}}^{m},$ which are
obtained by averaging over direction of vecotrs $\mathbf{k}$ and $\mathbf{k}%
^{\prime }$ and denoted $\Gamma _{\mathrm{kk}^{\prime }}^{m}$ (we use here a
short notation $\mathrm{k}=|\mathbf{k}|$ etc.)$,$ we obtain the system of
integral equations%
\begin{eqnarray}
\Gamma _{\mathrm{kk}^{\prime }}^{0} &=&\frac{2\pi \alpha _{r}}{k_{>}}%
K_{1}\left( \frac{\mathrm{k}_{<}}{\mathrm{k}_{>}}\right)  \notag \\
&&+\frac{\alpha _{r}}{4}\int \frac{d\mathrm{p}}{(\mathrm{p},\mathrm{k})_{>}}%
K_{1}\left( \frac{(\mathrm{p},\mathrm{k})_{<}}{(\mathrm{p},\mathrm{k})_{>}}%
\right) \left( \Gamma _{\mathrm{pk}^{\prime }}^{0}+\Gamma _{\mathrm{pk}%
^{\prime }}^{1}\right) ,  \notag \\
\Gamma _{\mathrm{kk}^{\prime }}^{1} &=&\frac{\alpha _{r}}{8}\int \frac{d%
\mathrm{p}}{(\mathrm{p},\mathrm{k})_{>}}K_{1}\left( \frac{(\mathrm{p},%
\mathrm{k})_{<}}{(\mathrm{p},\mathrm{k})_{>}}\right) \left( \Gamma _{\mathrm{%
pk}^{\prime }}^{0}+2\Gamma _{\mathrm{pk}^{\prime }}^{1}+\Gamma _{\mathrm{pk}%
^{\prime }}^{12}\right) ,  \notag \\
\Gamma _{\mathrm{kk}^{\prime }}^{12} &=&\frac{\alpha _{r}}{4}\int \frac{d%
\mathrm{p}}{(\mathrm{p},\mathrm{k})_{>}}K_{1}\left( \frac{(\mathrm{p},%
\mathrm{k})_{<}}{(\mathrm{p},\mathrm{k})_{>}}\right) \left( \Gamma _{\mathrm{%
pk}^{\prime }}^{12}+\Gamma _{\mathrm{pk}^{\prime }}^{1}\right) .
\label{Gamma_c}
\end{eqnarray}%
where $(\mathrm{p},\mathrm{k})_{>}=\max (\mathrm{p}\mathbf{,}\mathrm{k})$
and $(\mathrm{p},\mathrm{k})_{<}=\min (\mathrm{p}\mathbf{,}\mathrm{k}),$ $%
\mathrm{k}_{<}=(\mathrm{k},\mathrm{k}^{\prime })_{<},$ $\mathrm{k}_{>}=(%
\mathrm{k},\mathrm{k}^{\prime })_{>}$, $K_{1}(x)=(2/\pi )K(x),$ $K(x)$ is
the complete elliptic integral of the first kind. By considering the linear
combinations 
\begin{eqnarray}
\Gamma _{\mathrm{kk}^{\prime }}^{s} &=&\Gamma _{\mathrm{kk}^{\prime
}}^{0}+2\Gamma _{\mathrm{kk}^{\prime }}^{1}+\Gamma _{\mathrm{kk}^{\prime
}}^{12},  \notag \\
\Gamma _{\mathrm{kk}^{\prime }}^{d} &=&\Gamma _{\mathrm{kk}^{\prime
}}^{0}-2\Gamma _{\mathrm{kk}^{\prime }}^{1}+\Gamma _{\mathrm{kk}^{\prime
}}^{12},  \notag \\
\Gamma _{\mathrm{kk}^{\prime }}^{a} &=&\Gamma _{\mathrm{kk}^{\prime
}}^{0}-\Gamma _{\mathrm{kk}^{\prime }}^{12},  \label{linear}
\end{eqnarray}%
we decouple these equations to obtain%
\begin{eqnarray}
\Gamma _{\mathrm{kk}^{\prime }}^{s,a} &=&\frac{2\pi \alpha _{r}}{\mathrm{k}%
_{>}}K_{1}\left( \frac{\mathrm{k}_{<}}{\mathrm{k}_{>}}\right)  \notag \\
&&+\frac{\alpha _{r}r_{s,a}}{4}\int \frac{d\mathrm{p}}{(p,k)_{>}}K_{1}\left( 
\frac{(\mathrm{p},\mathrm{k})_{<}}{(\mathrm{p},\mathrm{k})_{>}}\right)
\Gamma _{\mathrm{pk}^{\prime }}^{s,a},  \notag \\
\Gamma _{\mathrm{kk}^{\prime }}^{d} &=&\frac{2\pi \alpha _{r}}{\mathrm{k}_{>}%
}K_{1}\left( \frac{\mathrm{k}_{<}}{\mathrm{k}_{>}}\right) .
\label{Gamma_dec}
\end{eqnarray}%
with $r_{s}=2,$ $r_{a}=1$. The solution to the resulting equations can be
searched in the form%
\begin{eqnarray}
\Gamma _{\mathrm{kk}^{\prime }}^{s,a} &=&\frac{1}{\mathrm{k}}g_{s,a}\left( 
\frac{\mathrm{k}^{\prime }}{\mathrm{k}}\right) ,\text{ }\mathrm{k}>\mathrm{k}%
^{\prime },  \label{Ans} \\
&=&\frac{1}{\mathrm{k}^{\prime }}g_{s,a}\left( \frac{\mathrm{k}}{\mathrm{k}%
^{\prime }}\right) ,\text{ }\mathrm{k}<\mathrm{k}^{\prime }.  \notag
\end{eqnarray}%
Often used approximation is to approximate $K_{1}(x)$ by its small $x$
asymptotic value, which is equal to $1.$ This corresponds to approximating 
\begin{equation}
1/|\mathbf{k}-\mathbf{k}^{\prime }|\approx 1/\mathrm{k}_{>}  \label{approx}
\end{equation}%
in the equation (\ref{EqGamma}). The considering s-wave component for the
vertex then coincides with the vertex itself, since the latter depends on
the absolute values of momenta only. The solution to the Eqs. (\ref%
{Gamma_dec}) reads (see Appendix A) 
\begin{equation}
g_{m}(x)=\frac{2\pi \alpha _{r}}{\sqrt{1-\alpha _{r}r_{m}}x^{\gamma _{m}}},
\label{gm}
\end{equation}%
where%
\begin{equation}
\gamma _{m}=\frac{1}{2}\left( 1-\sqrt{1-\alpha _{r}r_{m}}\right) .
\end{equation}%
Therefore, we find,%
\begin{eqnarray}
\Gamma _{\mathbf{kk}^{\prime }\mathbf{0}}^{0,12} &=&\frac{\pi \alpha _{r}}{2%
\sqrt{1-2\alpha _{r}}}\frac{1}{\mathrm{k}_{<}^{\gamma }\mathrm{k}%
_{>}^{1-\gamma }}\pm \frac{\pi \alpha _{r}}{\sqrt{1-\alpha _{r}}}\frac{1}{%
\mathrm{k}_{<}^{\gamma _{1}}\mathrm{k}_{>}^{1-\gamma _{1}}}+\frac{\pi \alpha
_{r}}{2\mathrm{k}_{>}},  \notag \\
\Gamma _{\mathbf{kk}^{\prime }\mathbf{0}}^{1} &=&\frac{\pi \alpha _{r}}{2%
\sqrt{1-2\alpha _{r}}}\frac{1}{\mathrm{k}_{<}^{\gamma }\mathrm{k}%
_{>}^{1-\gamma }}-\frac{\pi \alpha _{r}}{2\mathrm{k}_{>}},  \label{Gamma1}
\end{eqnarray}%
where%
\begin{equation}
\gamma =\gamma _{s},\gamma _{1}=\gamma _{a}.
\end{equation}%
The vertices diverge at $\alpha _{rc}=1/2$ which coincides with the
corresponding value, obtained from the analysis of the symmetry-broken phase
in Ref. \cite{Gusynin0,Khveschenko1,Murthy,Gusynin}. At $\alpha
_{r}\rightarrow \alpha _{rc}$ we find%
\begin{equation}
\Gamma _{\mathbf{kk}^{\prime }\mathbf{0}}^{0,1,12}\simeq \frac{\pi }{4\sqrt{%
1-2\alpha _{r}}}\frac{1}{(\mathrm{k}_{<}\mathrm{k}_{>})^{1/2}}.
\end{equation}%
More accurate analysis, considering angular dependence of the Coulomb
interaction, yields momentum dependences, which are qualitatively similar to
the Eq. (\ref{Gamma1}), with the exponents $\gamma ,$ $\gamma _{1}$,
determined by the Eq. (\ref{gammam1}) in the Appendix A. The exponent $%
\gamma $ coincides in this case with the result of Ref. \onlinecite{Gusynin}%
, obtained in the fall on the center problem. The critical interaction in
this case reduces to $\alpha _{rc}=0.46,$ see Ref. \onlinecite{Gusynin}.

The chiral susceptibility can be obtained by performing summation of the
ladder diagrams for the trianfular vertex $\Delta _{\mathbf{k}}$ via Eq. (%
\ref{Dp}), see Appendix A. To obtain the same result from the Eq. (\ref{Dp1}%
), we will also need subleading $1/\Lambda ^{\varkappa }$ corrections to the
vertex $\Gamma _{\mathbf{kk}^{\prime }\mathbf{0}}$. Evaluation of these
corrections can be performed in the way, similar to described above, and
yields%
\begin{eqnarray}
\Gamma _{\mathbf{kk}^{\prime }0}^{0}+2\Gamma _{\mathbf{kk}^{\prime
}0}^{1}+\Gamma _{\mathbf{kk}^{\prime }0}^{12} &=&\frac{2\pi \alpha _{r}}{%
\sqrt{1-2\alpha _{r}}}\left[ \frac{1}{\mathrm{k}_{>}^{1-\gamma }\mathrm{k}%
_{<}^{\gamma }}\right. \\
&-&\left. \frac{\gamma }{1-\gamma }\frac{1}{\Lambda ^{1-2\gamma }(\mathrm{kk}%
^{\prime })^{\gamma }}\right] .  \notag
\end{eqnarray}%
Integration of this vertex yields the triangular vertex (assuming $\Delta
_{0}=1$)%
\begin{eqnarray}
\Delta _{\mathbf{k}} &=&1+\frac{1}{2}\int\limits_{0}^{\Lambda }\left( \Gamma
_{\mathbf{kk}^{\prime }0}^{0}+2\Gamma _{\mathbf{kk}^{\prime }0}^{1}+\Gamma _{%
\mathbf{kk}^{\prime }0}^{12}\right) \frac{d\mathrm{k}^{\prime }}{2\pi } 
\notag \\
&=&\frac{\alpha _{r}}{2\gamma \sqrt{1-2\alpha _{r}}}\left[ 1-\frac{\gamma
^{2}}{(1-\gamma )^{2}}\right] \frac{\Lambda ^{\gamma }}{\mathrm{k}^{\gamma }}
\notag \\
&=&\frac{1}{1-\gamma }\frac{\Lambda ^{\gamma }}{\mathrm{k}^{\gamma }},
\label{delta}
\end{eqnarray}%
which is in agreement with the direct ladder summation (see Appendix A). We
see that in agreement with the scaling analysis of Ref. \cite{Gusynin} the
triangular vertex is non-singular at the chiral phase transition; the
singularity of the leading and subleading contributions to the vertex are
cancelled in $\Delta _{\mathbf{k}}$. The corresponding chiral susceptibility 
\begin{equation}
\chi =2\int\limits_{0}^{\Lambda }\frac{d\mathrm{k}}{2\pi }\Delta _{\mathbf{k}%
}=\frac{\Lambda }{\pi (1-\gamma )^{2}}=\frac{\Lambda }{2\pi \alpha _{r}}%
\frac{4\gamma }{1-\gamma }  \label{chi}
\end{equation}%
also does not diverge at the chiral phase transition.

\section{Functional renormalization-group analysis}

\subsection{General equations and parametrization of vertices}

The considered ladder approximation treats only a certain set of ladder
diagrams in the particle-hole channel. To consider the impact of other
diagramatic contributions, we apply in the present Section the functional
renormalization-group approach\cite{1PIRev,Salmhofer-paper,Salmhofer}. This
approach considers evolution of the electron interaction vertices due to
different scattering processes. Since the contribution of various channels
of electron interaction can compensate each other (see discussion below),
apriori the applicability of the ladder approximation, considered in
previous Section, corresponding to the account of particle-hole diagrams
only, is not clear.

To treat the possibility of chiral symmetry breaking within the functional
renormalization-group approach, we rewrite the action (\ref{L0}) in the form 
\begin{eqnarray}
L_{0} &=&-i\int \frac{d^{3}p}{(2\pi )^{3}}\ p_{\mu }\overline{\psi }%
_{ps}\gamma _{\mu }\psi _{ps}  \label{L0R} \\
&+&\frac{1}{2}\int \frac{d^{3}pd^{3}p^{\prime }d^{3}q}{(2\pi )^{9}}(%
\overline{\psi }_{ps}\gamma _{0}\psi _{p+q,s})\frac{2\pi \alpha }{|\mathbf{q}%
|}(\overline{\psi }_{p^{\prime }s^{\prime }}\gamma _{0}\psi _{p^{\prime
}-q,s^{\prime }})  \notag
\end{eqnarray}%
and consider the Wick-ordered functional RG flow equation\cite%
{Salmhofer-paper,Salmhofer} for the electron interaction vertex $%
V_{i_{1..4}}(k_{1},k_{2},k_{3})$ ($i_{1,2},k_{1,2}$ and $i_{3,4},k_{3,4}$
correspond to valley-sublattice indices and momenta/frequencies of incoming
and outgoing particles). The corresponding equation can be written
schematically as (see also Fig. 1), 
\begin{eqnarray}
\dot{V}_{\Lambda } &=&-V_{\Lambda }\circ \widetilde{\Pi }_{\Lambda ,\mathrm{%
pp}}\circ V_{\Lambda }  \label{fRG_Eq} \\
&&+V_{\Lambda }\circ \widetilde{\Pi }_{\Lambda ,\mathrm{ph}}\circ V_{\Lambda
}  \notag \\
&&+2V_{\Lambda }\ast \left( \widetilde{\Pi }_{\Lambda ,\mathrm{ph}}\circ -%
\widetilde{\Pi }_{\Lambda ,\mathrm{ph}}\ast \right) V_{\Lambda },  \notag
\end{eqnarray}%
where $\widetilde{\Pi }_{\Lambda ,\mathrm{pp,ph}%
}^{i_{1}i_{2};i_{3}i_{4}}(k,q)=D_{\Lambda }^{i_{1}i_{3}}(\mathbf{k},\nu
)F_{\Lambda }^{i_{2}i_{4}}(\mp \mathbf{k+q},\mp \nu +\omega )+F_{\Lambda
}^{i_{1}i_{3}}(\mathbf{k},\nu )D_{\Lambda }^{i_{2}i_{4}}(\mp \mathbf{k+q}%
,\mp \nu +\omega )$ is the fermionic bubble, constructed from the
Wick-ordering Green functions $D_{\Lambda }(\mathbf{k},\nu )$ and
single-scale propagators $F_{\Lambda }(\mathbf{k},\nu )$ (cf. Ref. %
\onlinecite{MyImp}), $\circ $ and $\ast $ stands for the summation over
momenta, frequencies, and quantum numbers according to the diagrammatic
rules (for explicit form of the flow equation see Appendix B, Eq. (\ref{Vdot}%
)).

The right-hand side of the Eq. (\ref{fRG_Eq}) contains several terms,
corresponding to the contribution of the particle-particle (pp),
particle-hole direct (ph) and particle-hole crossed (ph1, ph1$^{\prime }$)
processes (see also Fig. 1 and Appendix B). It is important to note that due
to the dispersion, which is odd in momentum, the contribution of the
particle-particle and particle-hole channels can compensate each other,
since for small external momenta the corresponding bubbles in the first two
diagrams in the right-hand side of Fig. 1 are equal in magnitude, but
opposite in sign (see Eq. (\ref{Pi})); last three diagrams also yield some
cancellation because of the sign change in the closed loop, and factor of
two in the spin summation in the last diagram.

These cancellations become transparent in the field-theoretical
renormalization-group approach \cite{Son,Son1,FostAleiner,Herbut2}, which
was applied previously to the same model. In particular, in this approach
the $\alpha ^{2}$ contributions to renormalization of short-range
interactions, which would drive the chiral instability similarly to quantum
chromodynamics \cite{Gies,Gies1}, are absent. At the same time,
field-theoretical renormalization group approach does not allow treating
singular momentum dependences, which are generated by the ladder diagrams in
the particle-hole channel, if the abovementioned cancellation between
particle-hole and particle-particle channel is weakly lifted by finite
external momenta. Therefore, we apply the functional renormalization-group
approach to the considering problem in this Section.

In the present study we use sharp momentum cutoff (cf. Refs. %
\onlinecite{KataninWick,MyImp}) 
\begin{equation}
\left. 
\begin{array}{c}
D_{\Lambda }(\mathbf{k},\varepsilon ) \\ 
F_{\Lambda }(\mathbf{k},\varepsilon )%
\end{array}%
\right\} =\overline{G}(\mathbf{k},\varepsilon )\left\{ 
\begin{array}{c}
\theta (\Lambda -|\mathbf{k|}) \\ 
\delta (\Lambda -|\mathbf{k}|)%
\end{array}%
\right. ,
\end{equation}%
where $\overline{G}=\left( \gamma _{0}\varepsilon +\mathrm{i}v_{F}\gamma
_{a}k_{a}-\Sigma (k)\right) ^{-1},$ neglect frequency dependence of the
vertices, and either neglect self-energy effects (taking $\Sigma (k)=0$) or
take the self-energy equal to its starting mean-field value, $\Sigma
(k)=(\alpha /4)v_{F}\gamma _{a}k_{a}\ln (\Lambda _{\mathrm{uv}}/|\mathbf{k}%
|),$ which coincides with the first-order perturbation theory result; $%
\Lambda _{\mathrm{uv}}$ is an ultraviolet cutoff of the theory. The latter
self-energy was recently shown to describe well the Fermi velocity
renormalization in free suspended graphene, see, e. g., Ref. \cite{Kopietz}.
The flow equations contain in both cases only the polarization bubble, which
is summed over fermionic Matsubara frequencies,%
\begin{eqnarray}
\widetilde{\Pi }_{\Lambda ,\mathrm{pp,ph}}(\mathbf{k},\mathbf{q})
&=&\sum\limits_{i\nu _{n}}\widetilde{\Pi }_{\Lambda ,\mathrm{pp,ph}%
}(k,q)_{\omega =0}  \notag \\
&=&\widetilde{\Pi }_{\mathrm{pp,ph}}(\mathbf{k},\mathbf{q})\delta (|\mathbf{k%
}|-\Lambda )\theta (\Lambda -|\mp \mathbf{k+q|})  \notag \\
&+&\widetilde{\Pi }_{\mathrm{pp,ph}}(\mp \mathbf{k+q},\pm \mathbf{q})\delta
(|\mp \mathbf{k+q}|-\Lambda )  \notag \\
&\times &\theta (\Lambda -|\mathbf{k|})^{{}}
\end{eqnarray}%
where $\widetilde{\Pi }_{\mathrm{pp,ph}}$ is given by the Eq. (\ref{Pi})
with $v_{F}(|\mathbf{k}|)=v_{F}$ or $v_{F}(|\mathbf{k}|)=v_{F}(1+(\alpha
/4)\ln (\Lambda _{\mathrm{uv}}/(|\mathbf{k}|))$. For $\Sigma =0$ the
functional renormalization group equations (\ref{fRG_Eq}) are analogous to
previously studied for the Hubbard model on the square- (Refs. %
\onlinecite{Metzner,HonerkampRice,Katanin,1PIRev}) and honeycomb (Ref. %
\onlinecite{Honerkamp}) lattices except that we use the continuum electronic
dispersion and concentrate on the effect of the long-range Coulomb
interaction. The use of the Wick-ordered approach allows us however to
include easily the renormalization of the Fermi velocity; it also
potentially allows to include in future effect impurities in this approach,
see the discussion in Ref. \onlinecite{MyImp}.

\begin{figure}[tb]
\includegraphics[width=7.7cm]{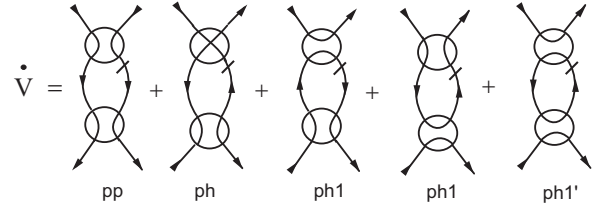} \vspace{-3mm}
\caption{Diagram representation of the flow equation (\protect\ref{fRG_Eq}).
Circles denote vertices $V_\Lambda$, solid lines without dash correspond to
Wick-ordering propagators $D_{\Lambda }$, while the lines with dash denote
the single-scale propagators $F_{\Lambda }$. The lines inside vertices show
direction of spin conservation.}
\end{figure}

We parametrize the vertices $V_{i_{1..4}}(\mathbf{k}_{1},\mathbf{k}_{2},%
\mathbf{k}_{3})$ in terms of the average momenta $\mathbf{k}_{\mathrm{pp}}=(%
\mathbf{k}_{1}-\mathbf{k}_{2})/2$, $\mathbf{k}_{\mathrm{ph}}=(\mathbf{k}_{2}+%
\mathbf{k}_{3})/2$, $\mathbf{k}_{\mathrm{ph1}}=(\mathbf{k}_{1}+\mathbf{k}%
_{3})/2$, and the momentum transfers $\mathbf{q}_{\mathrm{pp}}=\mathbf{k}%
_{1}+\mathbf{k}_{2}$, $\mathbf{q}_{\mathrm{ph}}=\mathbf{k}_{3}-\mathbf{k}%
_{2} $, and $\mathbf{q}_{\mathrm{ph1}}=\mathbf{k}_{1}-\mathbf{k}_{3}$ in the
particle-particle and particle-hole channels. For convenience we decompose
vertex functions into the contributions of the corresponding channels.
Written explicitly, our parametrization reads%
\begin{align}
V_{i_{1..4},\Lambda }(\mathbf{k}_{1},\mathbf{k}_{2},\mathbf{k}_{3})&
=V_{i_{1..4},\Lambda }^{\mathrm{pp}}(\mathbf{k}_{\mathrm{pp}},\mathbf{k}_{%
\mathrm{pp}}-\mathbf{q}_{\mathrm{ph1}},\mathbf{q}_{\mathrm{pp}})  \notag \\
& +V_{i_{1..4},\Lambda }^{\mathrm{ph}}(\mathbf{k}_{\mathrm{ph}}+\mathbf{q}_{%
\mathrm{ph1}},\mathbf{k}_{\mathrm{ph}},\mathbf{q}_{\mathrm{ph}})  \notag \\
& +V_{i_{1..4},\Lambda }^{\mathrm{ph1}}(\mathbf{k}_{\mathrm{ph1}},\mathbf{k}%
_{\mathrm{ph1}}-\mathbf{q}_{\mathrm{ph}},\mathbf{q}_{\mathrm{ph1}}).
\label{Dec}
\end{align}%
The first two arguments of functions $V_{i_{1..4}}^{\mathrm{pp,ph}}$
correspond to the average incoming and outgoing momentum, while the third
argument denotes the momentum transfer.

The functions $V_{i_{1..4},\Lambda }^{\mathrm{pp},\mathrm{ph}}$ of three
continuum $2$-component variables is hard to treat accurately numerically.
For $\mathrm{ph1}$ channel we pick out renormalized Coulomb interaction%
\begin{eqnarray}
&&V_{i_{1..4},\Lambda }^{\mathrm{ph1}}(\mathbf{k}_{\mathrm{ph1}},\mathbf{k}_{%
\mathrm{ph1}}-\mathbf{q}_{\mathrm{ph}},\mathbf{q}_{\mathrm{ph1}})  \notag \\
&=&\widetilde{V}_{i_{1..4},\Lambda }^{\mathrm{ph1}}(\mathbf{k}_{\mathrm{ph1}%
},\mathbf{k}_{\mathrm{ph1}}-\mathbf{q}_{\mathrm{ph}},\mathbf{q}_{\mathrm{ph1}%
})  \label{Vph1} \\
&+&\frac{2\pi \alpha }{\Pi _{\Lambda }(\mathbf{q}_{\mathrm{ph1}})}g_{\Lambda
,\mathbf{k}_{\mathrm{ph1}},\mathbf{q}_{\mathrm{ph1}}}^{i_{1}i_{3}}g_{\Lambda
,\mathbf{k}_{\mathrm{ph1}}-\mathbf{q}_{\mathrm{ph}},\mathbf{q}_{\mathrm{ph1}%
}}^{i_{2}i_{4}}  \notag
\end{eqnarray}%
by introducing the renormalized polarization $\Pi (\mathbf{q})$ and
electron-electric field vertex $g_{\Lambda ,\mathbf{k},\mathbf{q}}$
functions. We then approximate the functions $V_{i_{1..4},\Lambda }^{\mathrm{%
pp},\mathrm{ph}}$ and $\widetilde{V}_{i_{1..4},\Lambda }^{\mathrm{ph1}}$
neglecting their dependence on their third argument (i.e. transfer momenta $%
\mathbf{q}_{\mathrm{pp}}$, $\mathbf{q}_{\mathrm{ph}},$ or $\mathbf{q}_{%
\mathrm{ph1}}$), which are assumed to be zero in the actual calculation of
these quantities. The justification for this approximation is that at 
\textit{finite} $\mathbf{k},\mathbf{k}^{\prime }$ the dependence of these
functions on $\mathbf{q}_{\mathrm{pp,ph}}$ is non-singular, while there is
an essential singularity of the considering functions on the first two
momenta, $V_{i_{1..4},\Lambda }^{\mathrm{ph}}(\mathbf{k},\mathbf{k}^{\prime
},\mathbf{0})\mathbf{\sim }\max (\Lambda ,\mathbf{|k|,|k}^{\prime }\mathbf{|)%
}^{\gamma -1}$ $\times \max (\Lambda ,\min (\mathbf{|k|,|k}^{\prime }\mathbf{%
|))}^{-\gamma }$. We have also verified that taking into account the full
momentum dependence of $V_{i_{1..4},\Lambda }^{\mathrm{pp},\mathrm{ph}}$ and 
$\widetilde{V}_{i_{1..4},\Lambda }^{\mathrm{ph1}}$ yields only small
corrections to the obtained results.

To treat accurately the remaining momentum dependences, we follow the idea
of Ref. \cite{Salmhofer1}, expanding these functions in some basis. Since
the dependence on the absolute value of first two arguments is expected to
be singular (as follows from the analysis of Sect. II), we expand in Fourier
harmonics with respect to the angle of each of the two momenta. The
resulting flow equations for $V_{i_{1..4},\Lambda }^{\mathrm{pp},\mathrm{ph}%
} $, $\widetilde{V}_{i_{1..4},\Lambda }^{\mathrm{ph1}}$, $g_{\Lambda ,%
\mathbf{k},\mathbf{q}}$, and $\Pi _{\Lambda }(\mathbf{q})$ are presented in
the Appendix B.

\begin{figure}[tb]
\includegraphics[width=7.7cm]{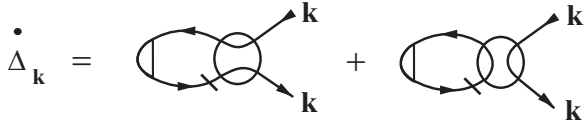} \vspace{-3mm}
\caption{Diagram representation of the flow equation for the triangular
vertex $\Delta _{\mathbf{k},\Lambda }$ (denoted as triangles in the
right-hand side). The first (both) term(s) in the right-hand side are
respectively included for the spin (charge) vertex. The notations are the
same, as in Fig. 1.}
\label{Fig:chi}
\end{figure}

To calculate the chiral spin or charge susceptibility, we first calculate
the flow of the triangular vertex $\Delta _{\mathbf{k},\Lambda }$ according
to the diagrams of Fig. \ref{Fig:chi}. The susceptibility is then obtained
straightforwardly by convolution of two triangular vertices with two Green
functions.

The flow is started at $\Lambda =\Lambda _{\mathrm{uv}}>1/v_{F},$ and
finished for some $\Lambda =\Lambda _{\min }\ll v_{F}.$ \ The bare value of
the vertex corresponds to the bare Coulomb interaction 
\begin{eqnarray}
g_{\Lambda _{\mathrm{uv}},\mathbf{k},\mathbf{q}}^{i_{1}i_{2}} &=&(\gamma
_{0})_{i_{1}i_{2}},  \notag \\
\Pi _{\Lambda _{\mathrm{uv}}}(\mathbf{q}) &=&|\mathbf{q}|,  \notag \\
V_{i_{1..4},\Lambda _{\mathrm{uv}}}^{\mathrm{pp}} &=&V_{i_{1..4},\Lambda _{%
\mathrm{uv}}}^{\mathrm{ph}}=\widetilde{V}_{i_{1..4},\Lambda _{\mathrm{uv}}}^{%
\mathrm{ph1}}=0,  \notag \\
\Delta^{i_1,i_2} _{\mathbf{k},\Lambda _{\mathrm{uv}}}&=&\delta_{i_1,i_2}.
\end{eqnarray}

\subsection{Results}

\subsubsection{Flow without the Fermi velocity renormalization}

First we neglect the self-energy effects by putting $\Sigma =0.$ In the
absence of self-energy effects, all the vertices have dimension $k^{-1},$
and can be expressed in terms of the scaling functions%
\begin{eqnarray}
V_{i_{1..4},m,n,\Lambda }^{\mathrm{pp,ph,ph}^{\prime }}(|\mathbf{k|},|%
\mathbf{k}^{\prime }|) &=&\frac{1}{|\mathbf{k|}}f_{i_{1}..i_{4},m,n}^{%
\mathrm{pp},\mathrm{ph,ph}^{\prime }}\left( \frac{|\mathbf{k}^{\prime }|}{|%
\mathbf{k}|},\frac{\Lambda }{|\mathbf{k}|},\frac{\Lambda _{\text{uv}}}{|%
\mathbf{k}|}\right) ,  \notag \\
\Pi _{\Lambda }(\mathbf{q}) &=&|\mathbf{q|}f^{\mathrm{\Pi }}\left( \frac{%
\Lambda }{|\mathbf{q}|},\frac{\Lambda _{\text{uv}}}{|\mathbf{q}|}\right) ,
\label{Scal} \\
g_{\Lambda ,\mathbf{k},\mathbf{q}}^{i_{1}i_{2}} &=&f_{i_{1},i_{2},m,n}^{%
\mathrm{g}}\left( \frac{|\mathbf{k}|}{|\mathbf{q}|},\frac{\Lambda }{|\mathbf{%
q}|},\frac{\Lambda _{\text{uv}}}{|\mathbf{q}|}\right)  \notag
\end{eqnarray}%
where $m,n$ numerates Fourier harmonics with respect to the directions of
momenta $\mathbf{k}$, $\mathbf{k}^{\prime },$ or $\mathbf{q}$. For
sufficiently large $\Lambda _{\text{uv}}\sim 1/v_{F}$ and small momenta, one
can ignore the last argument in these scaling functions.

The flow of $V^{\mathrm{ph}}(\mathbf{k,k}^{\prime })$ alone (with zero
interactions $V^{\mathrm{pp},\mathrm{ph1}^{\prime }}$) reproduces the
results of the ladder approach of Sect. II. We obtain the critical value $%
\alpha _{r}^{cr}=3.0/(2\pi )$ which is close to the value $\alpha
_{r}^{cr}=2.9/(2\pi )$, obtained in Refs. \cite{Khveschenko1,Murthy,Gusynin}%
, the difference is related to the discretization of momentum dependence of
the vertices. The flow of $\Pi (\mathbf{q})$ alone reproduces (in the end of
the flow) the renormalization of the dielectric constant by the static
polarization bubble, described by the Eq. (\ref{ar}),%
\begin{equation}
f^{\mathrm{\Pi }}\left( 0,\infty \right) =Z_{D}=1+\pi \alpha /2.
\end{equation}

\begin{figure}[tb]
\includegraphics[width=8.3cm]{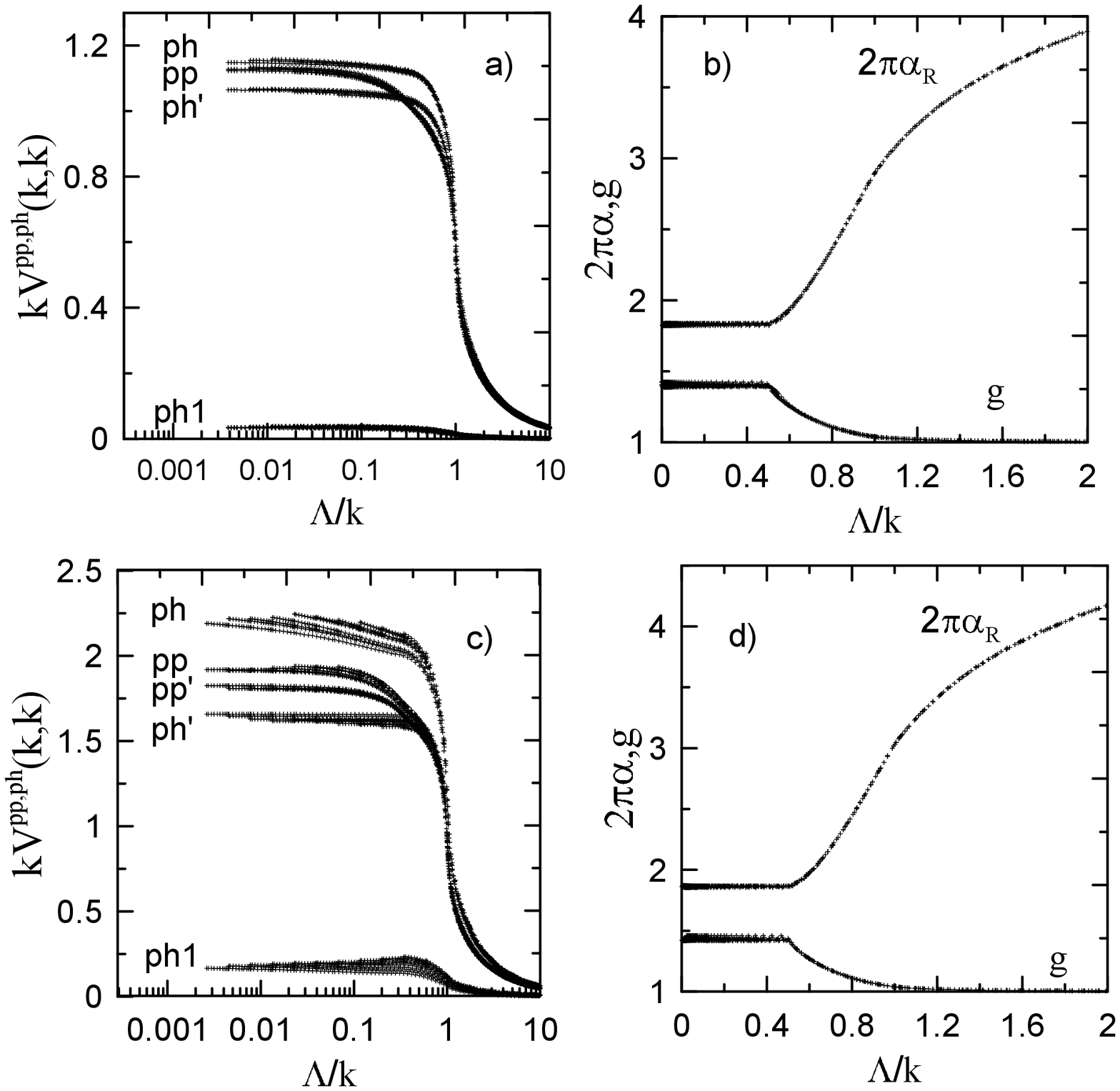} \vspace{-3mm}
\caption{Scale dependence of the absolute value of the vertex functions $|%
\mathbf{k|}V_{i_{1..4},m,n,\Lambda }^{\mathrm{pp,ph,ph1}}(|\mathbf{k|},|%
\mathbf{k}|)$ (a,c), the renormalized Coulomb interaction $\protect\alpha%
_{R,\Lambda}(\mathbf{q})$, and the electron-electric field interaction
vertex $g_{\Lambda,\mathbf{0},\mathbf{q}}^{1,1}$ (b,d) for $\protect\alpha %
=4/(2\protect\pi)$ (a,b) and $\protect\alpha =5.5/(2\protect\pi)$ (c,d).
Vertices pp,ph,ph1 correspond to the intrasublattice interaction $i_{1..4}=1$%
, while $\mathrm{pp}^{\prime}$,$\mathrm{ph^{\prime}}$ correspond to the
intersublattice interaction $i_1=i_3=1, i_2=i_4=2$.}
\label{Fig:Vf}
\end{figure}

\begin{figure}[tb]
\includegraphics[width=8cm]{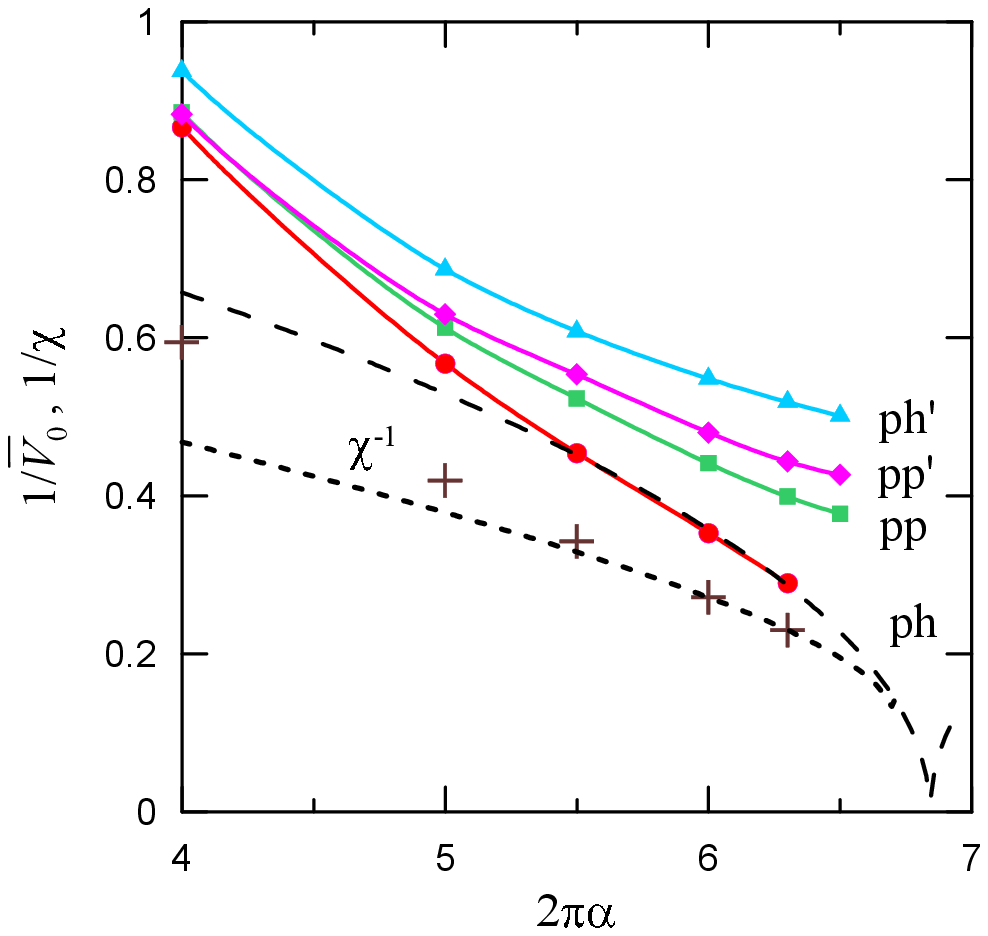} \vspace{-3mm}
\caption{(Color online) Interaction dependence of the absolute value of
inverse small momentum s-wave components of the dimensionless vertex
functions $\overline{V}_{0,i_{1..4},0,0}^{\mathrm{pp,ph}}$ and inverse
charge susceptibility in the end of the flow. Long-dashed line corresponds
to the fit of ph vertex function by the dependence $A_V/\protect\sqrt{%
\protect\alpha _{c,V}-\protect\alpha }$, long-dashed line corresponds to the
fit of the inverse susceptibility by $\protect\chi ^{-1}=A_{\protect\chi %
}(1+B_{\protect\chi }\protect\sqrt{\protect\alpha _{c,\protect\chi }-\protect%
\alpha })^{2}$. The other notations are the same as in Fig. \protect\ref%
{Fig:Vf}.}
\label{Fig:V0}
\end{figure}

\begin{figure}[tb]
\includegraphics[width=8.3cm]{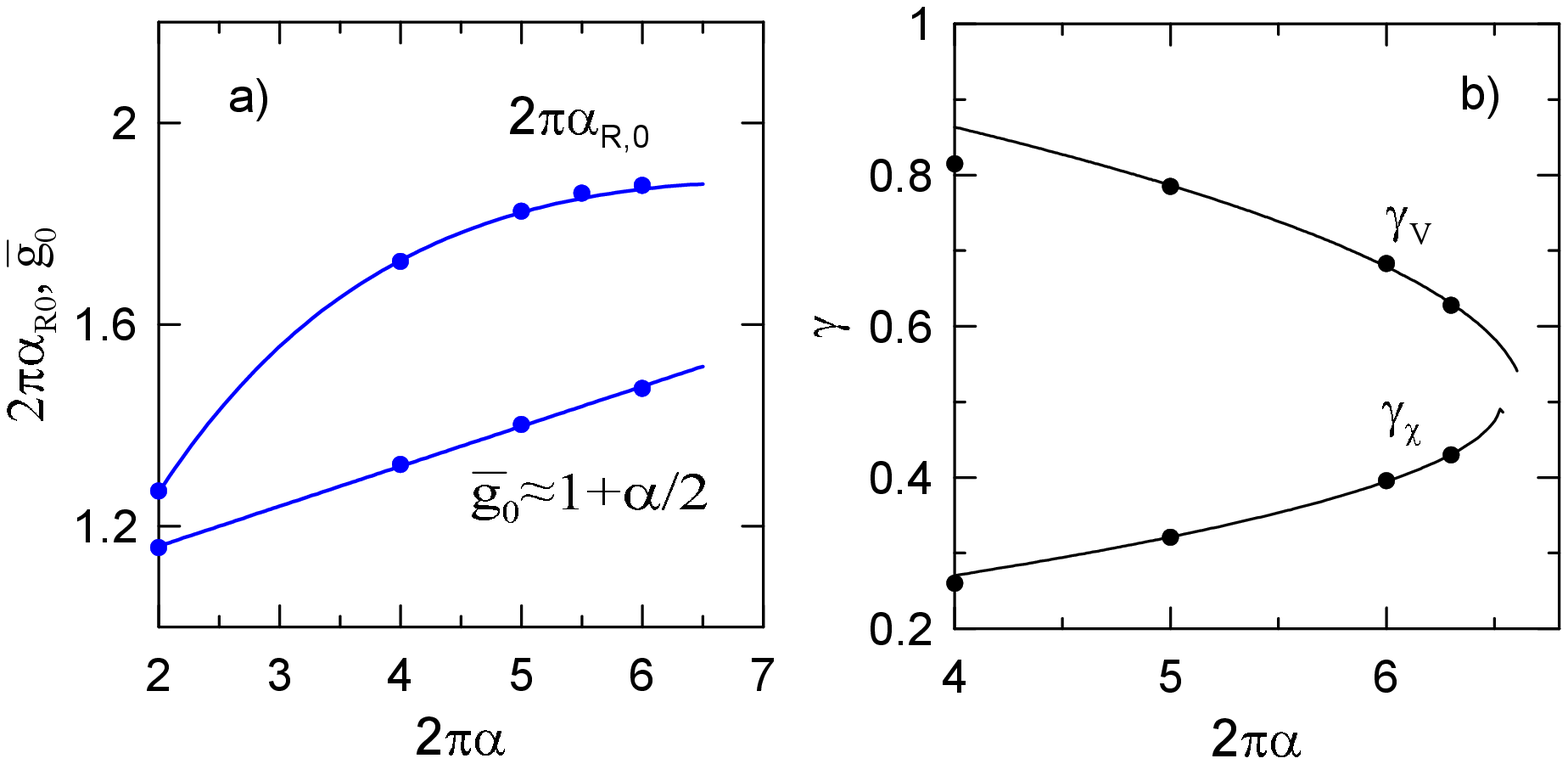} \vspace{-3mm}
\caption{(Color online) Interaction dependence of the renormalized Coulomb
interaction $\protect\alpha_{R,0}$, the electron-electric field interaction
vertex $\overline{g}_0^{1,1}$ (a), and the exponents of the momentum
dependence of the vertices $\protect\gamma_{V,\protect\chi}$ (b) is shown by
points. Upper solid line in (a) corresponds to the weak-coupling result (%
\protect\ref{Kotov_exp}) of Ref. \protect\cite{Sodemann}, while lower solid
line corresponds to the linear dependence $1+\protect\alpha /2$; solid line
in (b) is the result of the fit coording to the Eq. (\protect\ref{Defg})}
\label{Fig:AlphaGamma}
\end{figure}

In the present paper we consider however the combined contribution of all
the channels. To analyze the results, we plot dimensionless vertices $%
\overline{V}_{\Lambda ,i_{1}..i_{4},0,0}^{\mathrm{pp,ph,ph1}}(|\mathbf{k|})=|%
\mathbf{k|}V_{i_{1..4},0,0,\Lambda }^{\mathrm{pp,ph,ph1}}(|\mathbf{k|},|%
\mathbf{k}|)$ as functions of $\Lambda /|\mathbf{k}|,$ coupling constants $%
\alpha _{R,\Lambda }(|\mathbf{q|})=\alpha |\mathbf{q|}/\Pi _{\Lambda }(|%
\mathbf{q|}),$ and the vertex functions $g_{\Lambda ,\mathbf{0},\mathbf{q}%
}^{i_{1}i_{2}}$ as functions of $\Lambda /|\mathbf{q}|$ (see Fig. \ref%
{Fig:Vf}). We see, that the scaling forms (\ref{Scal}) are approximately
fulfilled. For small $\alpha \lesssim 5/(2\pi )$ we find that the
interactions, generated in the particle-particle and particle-hole channel
are close to each other in the absolute value, having the opposite signs,
and therefore, almost cancel each other. At the same time, for larger
interactions, this cancellation is lifted, and one of the coupling
constants, corresponding to the intrasublattice interaction in the
particle-hole channel, becomes bigger than the other interactions.

The interaction dependence of the inverse dimensionless quantities $%
\overline{V}_{0,i_{1}..i_{4},m,n}^{\mathrm{pp,ph}}=\lim_{x\rightarrow
0,y\rightarrow \infty }f_{i_{1}..i_{4},m,n}^{\mathrm{pp},\mathrm{ph}}\left(
1,x,y\right) $, corresponding to the limit $k\rightarrow 0,\Lambda
/k\rightarrow 0\ $of the coupling constants of Fig. \ref{Fig:Vf}, is shown
in Fig. \ref{Fig:V0}. We find that similarly to the ladder approach of Sect.
III, the leading intrasublattice component $\overline{V}_{0,1111}^{\mathrm{ph%
}}$ (which is equal to $|\mathbf{k|(}\Gamma _{\mathbf{kk0}}^{0}+\Gamma _{%
\mathbf{kk0}}^{12})-2\pi \alpha _{r}=$ $\pi \alpha _{r}(1/\sqrt{1-2\alpha
_{r}}-1)$ in the ladder case), increases faster than other components. By
fitting the obtained results with the dependence $A_{V}/\sqrt{\alpha
_{c,V}-\alpha }$ we find $\alpha _{c,V}=6.9/(2\pi ).$ From this fit we also
observe, that the ladder approximation with renormalized parameters is
applicable in fact only at $\alpha >5/(2\pi ),$ i.e. outside the region,
where particle-particle and particle-hole channels compensate each other.
The corresponding region below the critical interaction, which can be
associated with the critical regime, appears to be rather narrow.
Calculations of the chiral susceptibility shows that the spin and charge
susceptibility are equal in the considering model, since the second
contribution in the right-hand side of Fig. \ref{Fig:chi} vanishes.
Similarly to previous studies\cite{Herbut,Honerkamp}, additional short-range
interactions are expected to remove this degeneracy, yielding either spin or
charge order depending on the ratio of the on-site and nearest-neighbour
Coulomb interaction. Since the nearest-neighbour Coulomb interaction is
expected to be smaller than the on-site component, the spin-density wave is
expected to be more favourable, than the charge density wave. The
susceptibility, obtained within the present model, containing only
long-range interaction, is not singular near the chiral phase transition,
similarly to the ladder approximation. Its fit with the dependence $\chi
^{-1}=A_{\chi }(1+B_{\chi }\sqrt{\alpha _{c,\chi }-\alpha })^{2},$ similar
to the obtained in Eq. (\ref{chi}), yields $\alpha _{c,\chi }=6.7/(2\pi )$,
which is somewhat smaller $\alpha _{c,V}$. The small difference between $%
\alpha _{c,V}$ and $\alpha _{c,\chi }$ likely occurs because of the
narrowness of the region, where such fits are applicable.

We see that the critical coupling constants $\alpha _{c,V,\chi }$ are
approximately twice larger, than the constant $\alpha _{rc},$ obtained in
the ladder analysis. To understand the reason of this difference, we plot in
Fig. \ref{Fig:AlphaGamma}a the renormalized coupling constant $\alpha
_{R,0}=\alpha /Z_{D}^{\prime },$ where 
\begin{equation*}
Z_{D}^{\prime }=\lim_{q\rightarrow 0,\Lambda /q\rightarrow 0}\Pi _{\Lambda
}(|\mathbf{q|})/|\mathbf{q|}=\lim_{x\rightarrow 0,y\rightarrow \infty }f^{%
\mathrm{\Pi }}\left( x,y\right) ,
\end{equation*}%
and the vertex 
\begin{equation*}
\overline{g}_{0}^{i_{1}i_{2}}=\lim_{x,z\rightarrow 0,y\rightarrow \infty
}f_{i_{1},i_{2},m,n}^{\mathrm{g}}\left( z,x,y\right) .
\end{equation*}%
Due to vertex corrections, we obtain $Z_{D}^{\prime }<Z_{D}$, which agrees
with the observation of Ref. \cite{Kotov,Sodemann} that the vertex
corrections increase the effective dielectric constant. The obtained
behavior of $Z_{D}^{\prime }$ agrees well with the result of Ref. \cite%
{Sodemann}, 
\begin{equation}
Z_{D}^{\prime }=Z_{D}+0.78\alpha ^{2}+O(\alpha ^{3}),  \label{Kotov_exp}
\end{equation}%
although the latter is obtained in the weak-coupling limit (second order in $%
\alpha $). We also find, that the interaction dependence of the
intrasublattice component $\overline{g}_{0}^{11}$ is well fitted by $%
\overline{g}_{0}^{11}\simeq 1+\alpha /2$. This linear dependence of the
vertex (with the correct coefficient) can be obtained by comparing linear
and quadratic terms in Eq. (\ref{Kotov_exp}).

From Fig. \ref{Fig:AlphaGamma} we find that near chiral quantum phase
transition the coupling constant renormalization factor $\alpha
_{R,0}/\alpha \simeq 0.3,$ which is smaller than the ratio of the ladder and
fRG critical couplings $\alpha _{rc}/\alpha _{c}\simeq 0.43$. This
difference can be attributed to the vertex corrections. Although some vertex
corrections (yielding \textit{suppression }of the Coulomb interaction) are
already accounted by Eq. (\ref{Kotov_exp}), one should take into account,
that when constructing a particle-hole ladder, in the presence of the vertex
corrections each Coulomb line acquires a factor $g^{2}$. We find, that $(%
\overline{g}_{0}^{11})^{2}\simeq 2.25$ near the chiral phase transition.
This itself would yield two times \textit{decrease }of\textit{\ }the
critical interaction, which would be therefore equal $\alpha
_{rc}/0.3/2.25\simeq 0.68,$ somewhat smaller than the obtained value $\alpha
_{c}.$ The remaining difference can be explained by partial compensation
between particle-hole and the particle-particle channel near quantum phase
transition.

From Fig. \ref{Fig:Vf}c one can see that at $\alpha $ close to chiral phase
transition, the dimensionless coupling constants do not saturate yet in the
considered range of $\Lambda /q$ (the same behavior is observed for
susceptibilities, which are not shown). Considering smaller $\Lambda $
requires increasing number of discretization points, and not feasible
numerically. Apart from that, some deviations from scaling are observed,
which may be attributed to the discretization procedure. To determine
critical Coulomb interaction more accurately, we determine the exponents $%
\gamma _{V}$ and $\gamma _{\chi },$ defined by%
\begin{eqnarray}
V_{i_{1..4},0,0,\Lambda }^{\mathrm{ph}}(|\mathbf{k|},|\mathbf{k}^{\prime }|)
&=&\frac{B_{V}(k^{\prime })}{k^{\gamma _{V}}},\text{ }k^{\prime }<\Lambda <k,
\notag \\
\Delta _{\mathbf{k},\Lambda } &=&\frac{B_{\chi }(k)}{\Lambda ^{\gamma _{\chi
}}},\ \ k<\Lambda  \label{Defg}
\end{eqnarray}%
These exponents can be determined in the whole range $\Lambda \ll 1$ and do
not require achieving saturation of the coupling constants. In the ladder
approach one has $1-\gamma _{V}=$ $\gamma _{\chi }=\gamma .$ The results of
the renormalization-group calculation are shown in Fig. \ref{Fig:AlphaGamma}%
b. We see that the exponents $\gamma _{V},$ $\gamma _{\chi }$ also behave
similarly to the ladder approach for $2\pi \alpha >5,$ and show a
bifurcation point at $\alpha _{c}=1.048\pm 0.008.$ We consider this result
as an estimate for the critical coupling strength. This is much smaller than
the value $\alpha _{c}=1.62$ in the ladder approximation, and agrees well
with the results of Monte-Carlo analysis \cite{MC}.

\subsubsection{Flow with the Fermi velocity renormalization}

\begin{figure}[tb]
\includegraphics[width=8.3cm]{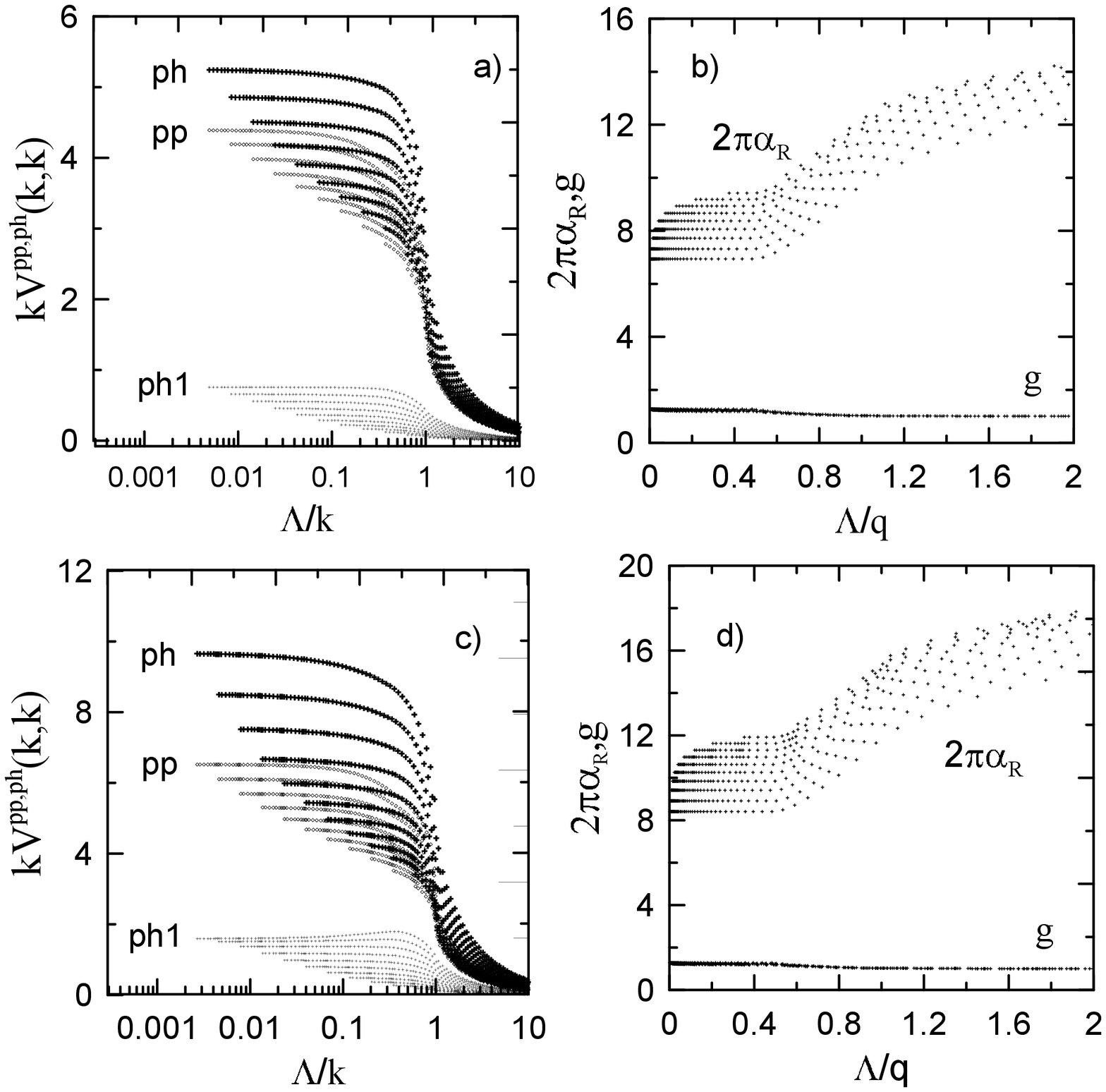} \vspace{-3mm}
\caption{Scale dependence of the absolute value of the vertex functions $|%
\mathbf{k|}V_{i_{1..4},m,n,\Lambda }^{\mathrm{pp,ph,ph1}}(|\mathbf{k|},|%
\mathbf{k}|)$ (a,c), the renormalized Coulomb interaction $\protect\alpha%
_{R,\Lambda}(\mathbf{q})$, and the electron-electric field interaction
vertex $g_{\Lambda,\mathbf{0},\mathbf{q}}^{1,1}$ (b,d) for $\protect\alpha %
=16/(2\protect\pi)$ (a,b) and $\protect\alpha =20/(2\protect\pi)$ (c,d) with
account of the Fermi velocity renormalization. The notations are the same,
as in Fig. \protect\ref{Fig:Vf}.}
\label{Fig:Vf1}
\end{figure}

\begin{figure}[tb]
\includegraphics[width=8cm]{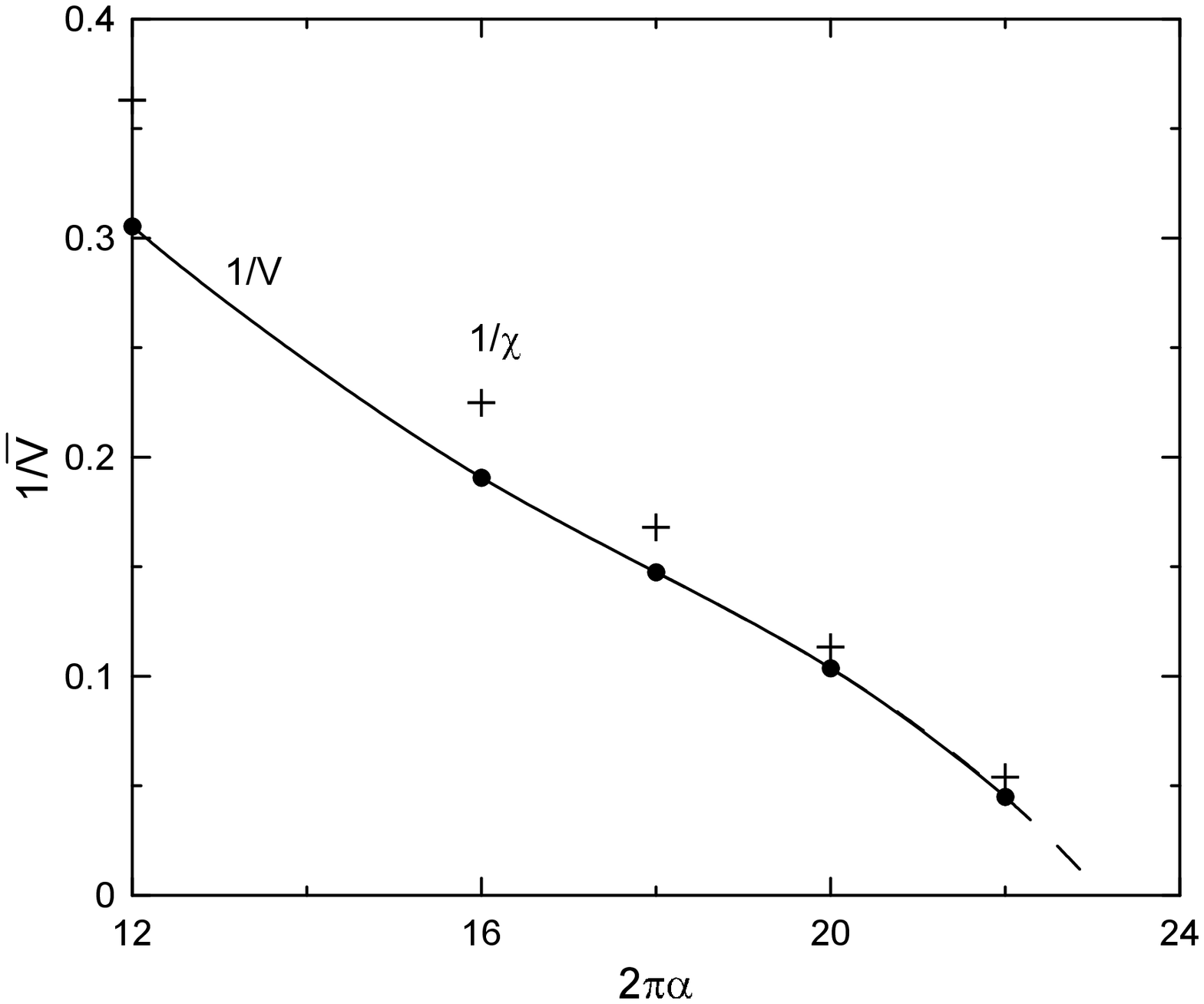} \vspace{-3mm}
\caption{Interaction dependence of the absolute value of inverse small
momentum s-wave components of the dimensionless vertex functions $\overline{V%
}_{0,\{1,1,1,1\},0,0}^{\mathrm{ph}}$ (line and points) and inverse charge
susceptibility (crosses) in the end of the flow with account of the
renormalization of the Fermi velocity. Long-dashed line corresponds to the
extrapolation of interaction dependence of the inverse vertex.}
\label{Fig:Vv0}
\end{figure}

To model the effect of Fermi velocity renormalization, we put $\Sigma
(k)=(\alpha /4)v_{F}\gamma _{a}k_{a}\ln (\Lambda _{uv}/|\mathbf{k|})$. The
scale dependences of the vertex functions are shown in Fig. \ref{Fig:Vf1}.
As one can expect, the universality of the scaling functions is violated by
renormalization of the Fermi velocity. Due to relatively large Coulomb
interaction, the contributions of the particle-particle and particle-hole
channels are split, such that the above discussed compensation of the two
contributions is less pronouced. The electron-electric field vertex
renormalization achieves values $\gamma \simeq 1.5,$ which are comparable to
those obtained in the previous subsection without Fermi velocity
renormalization (although for larger interaction strength), strongly
suppressing the critical interactions. The dependence of the interaction in
the end of the flow on the coupling constants is shown in Fig. \ref{Fig:Vv0}%
. We obtain the critical coupling constant $\alpha _{c}^{v}=3.7,$ while
without the vertex corrections in the Bethe-Salpeter approach the chiral
instability does not occur in the static approximation with the Fermi
velocity renormalization\cite{Popovici}. Apart from the vertex corrections,
discussed in previous Section, the reason of getting finite and not large
critical interaction is in the renormalization of Fermi velocity also when
calculating the polarization bubble. This effect was not accounted in Ref. %
\onlinecite{Popovici} and reduces the screening of the interaction,
decreasing therefore the effective interaction. To see more explicitly the
effect of changing screening in the presence of the renormalization of the
Fermi velocity, we switch off the renormalization of the Fermi velocity in
the polarization bubble $\Pi _{\Lambda }(\mathbf{q})$ and obtain $\alpha
_{c}^{v,0}=4.6$ which is somewhat higher than the $\alpha _{c}^{v},$ but yet
finite and even considerably smaller than the result of Ref. %
\onlinecite{Popovici} with dynamic effects included. Therefore, we conclude
that even with the Fermi velocity renormalization, vertex corrections yield
substantial decrease of the critical interaction. The obtained value $\alpha
_{c}^{v}=3.7$ is expected to be further reduced by the effects of dynamic
screening of Coulomb interaction, which may yield the critical Coulomb
interactrion comparable to the one in suspended graphene.

\section{Conclusion}

In the present paper we have analyzed the results of the Bethe-Salpeter
(ladder) approximation and the functional renormalization-group approach,
which accounts for all the channels of electron interaction to describe
chiral phase transition in a system of Dirac electrons. We have shown, that
without the Fermi velocity renormalization, at sufficiently small coupling
constants $\alpha <5/(2\pi )$ the particle-particle and particle-hole
channels partly compensate each other. On the other hand, for $\alpha
>5/(2\pi )$ the compensation is not present, and the particle-hole channel
dominates. This yields chiral phase transition at $\alpha _{c}=1.05,$ which
properties are rather similar to those, obtained in the Schwinger-Dyson or
Bethe-Salpeter (ladder) analysis. The vertex corrections enhance the
tendency towards chiral symmetry breaking and compensate the effect of
partial cancellation between the particle-particle and particle-hole
channels.

With accounting the Fermi velocity renormalization, the abovementioned
compensation is weakened, and therefore vertex corrections are expected to
have even stronger effect on the critical interaction sterngth. In
particular, in the static approximation we obtain $\alpha _{c}^{v}=3.7$,
while this value is expected to be further reduced by the effects of dynamic
screening of Coulomb interaction, yielding the results for the critical
Coulomb interaction smaller than previous estimates. Whether it remains
above or below the experimental $\alpha =2.2$ and how much it differs from
the results of Monte-Carlo analysis, requires additional studies. As also
described in the introduction, the effect of the other bands, which add
short-range interactions on the top of the long-range Coulomb tail, can be
also important. Their analysis is beyond of the scope of the present work,
and is interesting to be performed in the future. Finally, the effect of
impurities, either neutral, or charged, in the presence of the interelectron
Coulomb interaction, has to be analyzed.

Acknowledgements. The author is grateful to M. Scherer, F. Assaad, M. I.
Katsnelson, M. Titov, S. Friedrich, and C. Wetterich for discussions. The
work is supported by FASO Russian Federation (theme "Electron" No.
01201463326), grant of the Dynasty foundation, and Act 211 Government of the
Russian Federation, contract 02.A03.21.0006. The calculations are performed
using computer cluster Uran of Ural branch RAS.

\section*{Appendix. A. Solution of the Bethe-Salpeter equation for the vertex%
}

Substituting the ansatz (\ref{Anz}) into the Eq. (\ref{EqGamma}) we find 
\begin{eqnarray}
&&\left( \Gamma _{\mathbf{kk}^{\prime }\mathbf{0}}^{0}-\frac{2\pi \alpha }{|%
\mathbf{k}-\mathbf{k}^{\prime }|}\right) \gamma _{0}\otimes \gamma
_{0}+\Gamma _{\mathbf{kk}^{\prime }\mathbf{0}}^{1}\gamma _{a}\otimes \gamma
_{a}  \label{EqG} \\
&&+\Gamma _{\mathbf{kk}^{\prime }\mathbf{0}}^{12}i\gamma _{0}\gamma
_{1}\gamma _{2}\otimes i\gamma _{0}\gamma _{0}\gamma _{1}\gamma _{2}  \notag
\\
&=&\frac{\alpha _{r}}{4}\int \frac{dpd\theta _{p}}{2\pi }\frac{1}{|\mathbf{p}%
-\mathbf{k|}}\left[ I\otimes I-\frac{p_{a}p_{b}}{|\mathbf{p}|^{2}}\gamma
_{0}\gamma _{a}\otimes \gamma _{0}\gamma _{a}\right]  \notag \\
&\times &\left[ \Gamma _{\mathbf{pk}^{\prime }\mathbf{0}}^{0}\gamma
_{0}\otimes \gamma _{0}+\Gamma _{\mathbf{pk}^{\prime }\mathbf{0}}^{1}\gamma
_{b}\otimes \gamma _{b}\right.  \notag \\
&&+\left. \Gamma _{\mathbf{pk}^{\prime }\mathbf{0}}^{12}i\gamma _{0}\gamma
_{1}\gamma _{2}\otimes i\gamma _{0}\gamma _{1}\gamma _{2}\right] .  \notag
\end{eqnarray}%
In the following we consider the "s-wave" component of the vertex, which is
obtained by averaging over the directions of $\mathbf{k}$ and $\mathbf{k}%
^{\prime }.$ For this component, which we refer in the following as $\Gamma
_{kk^{\prime }}^{m},$ we obtain, 
\begin{eqnarray}
&&\left[ \Gamma _{kk^{\prime }}^{0}-\frac{2\pi \alpha }{k_{>}}K_{1}\left( 
\frac{k_{<}}{k_{>}}\right) \right] \gamma _{0}\otimes \gamma _{0}+\Gamma
_{kk^{\prime }}^{1}\gamma _{a}\otimes \gamma _{a}  \notag \\
&&+\Gamma _{kk^{\prime }}^{12}i\gamma _{0}\gamma _{1}\gamma _{2}\otimes
i\gamma _{0}\gamma _{0}\gamma _{1}\gamma _{2} \\
&=&\frac{\alpha _{r}}{4}\int \frac{dp}{(p,k)_{>}}K_{1}\left( \frac{(p,k)_{<}%
}{(p,k)_{>}}\right) \left[ I\otimes I-\frac{1}{2}\gamma _{0}\gamma
_{a}\otimes \gamma _{0}\gamma _{a}\right]  \notag \\
&\times &\left[ \Gamma _{pk^{\prime }}^{0}\gamma _{0}\otimes \gamma
_{0}+\Gamma _{pk^{\prime }}^{1}\gamma _{b}\otimes \gamma _{b}+\Gamma
_{pk^{\prime }}^{12}i\gamma _{0}\gamma _{1}\gamma _{2}\otimes i\gamma
_{0}\gamma _{1}\gamma _{2}\right] ,  \notag
\end{eqnarray}%
where $K_{1}(x)=(2/\pi )K(x),$ $K(x)$ is the complete elliptic integral of
the first kind, $k_{<}=\min (k,k^{\prime })$, $k_{>}=\max (k,k^{\prime }),$ $%
(p,k)_{>}=\max (p,k),$ $(p,k)_{<}=\min (p,k)$. Performing matrix
multiplication, we obtain Eq. (\ref{Gamma_c}) of the main text.

With the anzatz (\ref{Ans}) for the linear combinations (\ref{linear}), we
obtain for the functions $g_{m}(x)$ with $m=s,a$ the equations%
\begin{eqnarray}
\frac{1}{k}g_{m}(\frac{k^{\prime }}{k}) &=&\frac{2\pi \alpha _{r}}{k}K\left( 
\frac{k^{\prime }}{k}\right) +\frac{\alpha _{r}r_{m}}{4kk^{\prime }}%
\int_{0}^{k^{\prime }}dpg_{m}\left( \frac{p}{k^{\prime }}\right) K_{1}\left( 
\frac{p}{k}\right)  \notag \\
&&+\frac{\alpha _{r}r_{m}}{4k}\int_{k^{\prime }}^{k}\frac{dp}{p}g_{m}\left( 
\frac{k^{\prime }}{p}\right) K_{1}\left( \frac{p}{k}\right)  \notag \\
&&+\frac{\alpha _{r}}{4}\int_{k}^{\Lambda }\frac{dp}{p^{2}}g_{m}\left( \frac{%
k^{\prime }}{p}\right) K_{1}\left( \frac{k}{p}\right) .
\end{eqnarray}%
Supposing the last integral is convergent at $\Lambda \rightarrow \infty $,
these equations can be rewritten as 
\begin{eqnarray}
g_{m}(x) &=&2\pi \alpha _{r}K_{1}\left( x\right) +\frac{\alpha _{r}r_{m}}{4}%
\int_{0}^{1}d\widetilde{p}g_{m}(\widetilde{p})K_{1}\left( \widetilde{p}%
x\right)  \notag \\
&&+\frac{\alpha _{r}r_{m}}{4}\int_{x}^{1}\frac{d\widetilde{p}}{\widetilde{p}}%
g_{m}(\widetilde{p})K_{1}\left( \frac{x}{\widetilde{p}}\right)  \notag \\
&&+\frac{\alpha _{r}r_{m}}{4x}\int_{0}^{x}d\widetilde{p}g_{m}(\widetilde{p}%
)K_{1}\left( \frac{\widetilde{p}}{x}\right) .  \label{gEq}
\end{eqnarray}%
The obtained integral equation can be then transformed to the differential
one, 
\begin{equation}
x[xg_{m}(x)]^{\prime \prime }=-\frac{\alpha _{r}r_{m}}{4}g_{m}(x)
\end{equation}%
with 
\begin{equation}
g_{m}(1)-2\pi \alpha =-2g_{m}^{\prime }(1)=\frac{\alpha _{r}r_{m}}{2}%
\int_{0}^{1}dxg_{m}(x).
\end{equation}%
The solution to the obtained equations has the form of the Eq. (\ref{gm}),
of the main text, which yields 
\begin{eqnarray}
k_{>}\Gamma _{kk^{\prime }}^{1} &=&\frac{\pi \alpha _{r}}{2\sqrt{1-2\alpha
_{r}}}\left( \frac{k_{>}}{k_{<}}\right) ^{\gamma }-\frac{\pi \alpha _{r}}{2},
\label{res} \\
k_{>}\Gamma _{kk^{\prime }}^{0,12} &=&\frac{\pi \alpha _{r}}{2\sqrt{%
1-2\alpha _{r}}}\left( \frac{k_{>}}{k_{<}}\right) ^{\gamma }  \notag \\
&\pm &\frac{\pi \alpha _{r}}{\sqrt{1-\alpha _{r}}}\left( \frac{k_{>}}{k_{<}}%
\right) ^{\gamma _{a}}+\frac{\pi \alpha _{r}}{2},  \notag
\end{eqnarray}%
where we have denoted $\gamma =\gamma _{s}=\frac{1}{2}\left( 1-\sqrt{%
1-2\alpha _{r}}\right) .$

The solution of the equations (\ref{gEq}) beyond the approximation (\ref%
{approx}) can be obtained analytically only for $x=k^{\prime }/k\ll 1.$
Assuming again $g_{m}(x\ll 1)=A_{m}x^{-\gamma _{m}}$ we find: 
\begin{eqnarray}
A_{m}x^{-\gamma _{m}} &=&2\pi \alpha +\frac{\alpha r_{m}}{4}\int_{0}^{1}d%
\widetilde{p}g_{m}(\widetilde{p}) \\
&&+\frac{\alpha r_{m}}{4}\int_{x}^{1}\frac{d\widetilde{p}}{\widetilde{p}}%
g_{m}(\widetilde{p})K_{1}\left( \frac{x}{\widetilde{p}}\right)  \notag \\
&&+\frac{\alpha r_{m}A_{m}}{4x}\int_{0}^{x}d\widetilde{p}\widetilde{p}%
^{-\gamma _{m}}K_{1}\left( \frac{\widetilde{p}}{x}\right) .  \notag
\end{eqnarray}%
To simplify the second integral, we introduce the constant $C$ such that $%
x\ll C\ll 1.$ Splitting the regions of integration over $\widetilde{p}$ by
the constant $C$, we find 
\begin{eqnarray}
A_{m}x^{-\gamma _{m}} &=&2\pi \alpha _{r}+\frac{\alpha _{r}r_{m}}{4}%
\int_{0}^{1}d\widetilde{p}g_{m}(\widetilde{p}) \\
&&+\frac{\alpha _{r}r_{m}A_{m}}{4}\int_{x}^{C}\frac{d\widetilde{p}}{%
\widetilde{p}^{1+\gamma _{m}}}K_{1}\left( \frac{x}{\widetilde{p}}\right) 
\notag \\
&&+\frac{\alpha _{r}r_{m}}{4}\int_{C}^{1}\frac{d\widetilde{p}}{\widetilde{p}}%
g_{m}(\widetilde{p})  \notag \\
&&+\frac{\alpha _{r}r_{m}A_{m}}{4x}\int_{0}^{x}d\widetilde{p}\widetilde{p}%
^{-\gamma _{m}}K_{1}\left( \frac{\widetilde{p}}{x}\right)  \notag
\end{eqnarray}%
\begin{eqnarray*}
&=&2\pi \alpha _{r}+\frac{\alpha _{r}r_{m}}{4}\int_{0}^{1}d\widetilde{p}%
g_{m}(\widetilde{p})-\frac{\alpha _{r}r_{m}A_{m}}{4\gamma _{m}} \\
&&+\frac{\alpha _{r}r_{m}}{4}\int_{0}^{1}\frac{d\widetilde{p}}{\widetilde{p}}%
\left[ g_{m}(\widetilde{p})-\frac{A_{m}}{\widetilde{p}^{\gamma _{m}}}\right]
\\
&&+\frac{\alpha _{r}r_{m}A_{m}}{4x^{\gamma _{m}}}\left[ \int_{0}^{1}dyy^{-%
\gamma _{m}}K_{1}\left( y\right) \right. \\
&&\left. +\int_{1}^{\infty }\frac{dy}{y^{1+\gamma _{m}}}K_{1}\left( \frac{1}{%
y}\right) \right] .
\end{eqnarray*}%
The resulting equation for $\gamma _{m}$ reads%
\begin{equation}
1=\frac{\alpha _{r}r_{m}}{4}\int_{0}^{1}dy\left( y^{\gamma
_{m}-1}+y^{-\gamma _{m}}\right) K\left( y\right) ;  \label{gammam1}
\end{equation}%
the result for the exponent $\gamma =\gamma _{s}$ coincides with that
obtained in the fall on the center problem\cite{Gusynin}. For the constant $%
A_{m}$ we obtain the equation 
\begin{eqnarray}
&&\frac{\alpha _{r}r_{m}A_{m}}{4\gamma _{m}}-\frac{\alpha _{r}r_{m}}{4}%
\int_{0}^{1}d\widetilde{p}g_{m}(\widetilde{p})  \notag \\
&&-\frac{\alpha _{r}r_{m}}{4}\int_{0}^{1}\frac{d\widetilde{p}}{\widetilde{p}}%
\left[ g_{m}(\widetilde{p})-\frac{A_{m}}{p^{\gamma _{m}}}\right] 
\begin{array}{c}
=%
\end{array}%
2\pi \alpha
\end{eqnarray}%
which can be solved numerically. Qualitatively, the solution do not differ
from the simplified result (\ref{res}).

Similar consideration can be performed for the triangular vertex, considered
in Ref. \cite{Gusynin1}, cf. also main text. The corresponding equation for
the "s-wave" component of the vertex reads 
\begin{equation}
\Delta _{p}=\Delta _{0}+\frac{\alpha _{r}}{2}\int\limits_{0}^{\Lambda }\frac{%
dk}{(k,p)_{>}}K_{1}\left( \frac{(k,p)_{<}}{(k,p)_{>}}\right) \Delta _{k},
\end{equation}%
where $\Delta _{0}$ is the bare value of the triangular vertex. Substituting 
$\Delta _{p}=Ap^{-\gamma }$ for $p\ll \Lambda $ and introducing $p\ll C\ll
\Lambda ,$ we obtain for $\gamma $ the equation 
\begin{eqnarray}
Ap^{-\gamma } &=&\Delta _{0}+\frac{\alpha _{r}A}{2p}\int\limits_{0}^{p}\frac{%
dk}{k^{\gamma }}K_{1}\left( \frac{k}{p}\right) \\
&&+\frac{\alpha _{r}A}{2}\int\limits_{p}^{C}\frac{dk}{k^{1+\gamma }}%
K_{1}\left( \frac{p}{k}\right) +\frac{\alpha _{r}}{2}\int\limits_{C}^{%
\Lambda }\frac{dk}{k}\Delta _{k}  \notag \\
&=&\Delta _{0}+\frac{\alpha _{r}A}{2p^{\gamma }}\left[ \int\limits_{0}^{1}%
\frac{dy}{y^{\gamma }}K_{1}\left( y\right) +\int\limits_{1}^{\infty }\frac{dk%
}{y^{1+\gamma }}K_{1}\left( \frac{1}{y}\right) \right]  \notag \\
&&+\frac{\alpha _{r}}{2}\int\limits_{0}^{\Lambda }\frac{dk}{k}\left( \Delta
_{k}-\frac{A}{k^{\gamma }}\right) -\frac{\alpha _{r}A\Lambda ^{\gamma }}{%
2\gamma }.  \notag
\end{eqnarray}%
From this we find again $\gamma =\gamma _{s},$ determined above, and the
constant $A$ is determined by 
\begin{equation}
\frac{\alpha _{r}A\Lambda ^{\gamma }}{2\gamma }-\frac{\alpha _{r}}{2}%
\int\limits_{0}^{\Lambda }\frac{dk}{k}\left( \Delta _{k}-\frac{A}{k^{\gamma }%
}\right) =\Delta _{0}.
\end{equation}%
With the approximation (\ref{approx}) this reduces to the result (\ref{delta}%
) of the main text.

\section*{Appendix. B. Functional renormalization group equations}

In this Appendix we present analytical form of the renormalization-group
equations, shown in Fig. 1 of the paper. The equation for the vertices $%
V_{i_{1..4}}^{m}(\mathbf{k}_{1},\mathbf{k}_{2},\mathbf{k}_{3})$ in the
static approximation reads:%
\begin{align}
& \dot{V}_{i_{1..4},\Lambda }(\mathbf{k}_{1},\mathbf{k}_{2},\mathbf{k}%
_{3})=-\sum\limits_{\mathbf{k,}i_{1..4}^{\prime }}  \notag \\
& \left\{ V_{i_{1}i_{4}^{\prime }i_{1}^{\prime }i_{4},\Lambda }(\mathbf{k}%
_{1},\mathbf{k-q}_{\mathrm{ph}},\mathbf{k})V_{i_{3}^{\prime
}i_{2}i_{3}i_{2}^{\prime },\Lambda }(\mathbf{k},\mathbf{k}_{2},\mathbf{k}%
_{3})\right.  \notag \\
& \times \widetilde{\Pi }_{\Lambda ,\mathrm{ph}}^{i_{1}^{\prime
}i_{2}^{\prime };i_{3}^{\prime }i_{4}^{\prime }}(\mathbf{k},\mathbf{q}_{%
\mathrm{ph}})  \notag \\
& +V_{i_{1}i_{2}i_{1}^{\prime }i_{2}^{\prime },\Lambda }(\mathbf{k}_{1},%
\mathbf{k}_{2},\mathbf{k})V_{i_{3}^{\prime }i_{4}^{\prime
}i_{3}i_{4},\Lambda }(\mathbf{k},-\mathbf{k}+\mathbf{q}_{\mathrm{pp}},%
\mathbf{k}_{3})  \notag \\
& \times \widetilde{\Pi }_{\Lambda ,\mathrm{pp}}^{i_{1}^{\prime
}i_{2}^{\prime };i_{3}^{\prime }i_{4}^{\prime }}(\mathbf{k},\mathbf{q}_{%
\mathrm{pp}})+  \notag \\
& \left[ V_{i_{1}i_{1}^{\prime }i_{3}^{\prime }i_{3},\Lambda }(\mathbf{k}%
_{1},\mathbf{k}-\mathbf{q}_{\mathrm{ph1}},\mathbf{k})V_{i_{4}^{\prime
}i_{2}i_{2}^{\prime }i_{4},\Lambda }(\mathbf{k},\mathbf{k}_{2},\mathbf{k-q}_{%
\mathrm{ph}})\right.  \notag \\
& +V_{i_{1}i_{1}^{\prime }i_{3}i_{3}^{\prime },\Lambda }(\mathbf{k}_{1},%
\mathbf{k}-\mathbf{q}_{\mathrm{ph1}},\mathbf{k}_{3})V_{i_{4}^{\prime
}i_{2}i_{4}i_{2}^{\prime },\Lambda }(\mathbf{k},\mathbf{k}_{2},\mathbf{k}%
_{2}+\mathbf{q}_{\mathrm{ph1}})  \notag \\
& \left. -2V_{i_{1}i_{1}^{\prime }i_{3}i_{3}^{\prime },\Lambda }(\mathbf{k}%
_{1},\mathbf{k}-\mathbf{q}_{\mathrm{ph1}},\mathbf{k}_{3})V_{i_{4}^{\prime
}i_{2}i_{2}^{\prime }i_{4},\Lambda }(\mathbf{k},\mathbf{k}_{2},\mathbf{k-q}_{%
\mathrm{ph}})\right]  \notag \\
& \left. \times \widetilde{\Pi }_{\Lambda ,\mathrm{ph}}^{i_{2}^{\prime
}i_{3}^{\prime };i_{1}^{\prime }i_{4}^{\prime }}(\mathbf{k},\mathbf{q}_{%
\mathrm{ph1}})\right\} .  \label{Vdot}
\end{align}%
where $\mathbf{k}_{\mathrm{pp}}=(\mathbf{k}_{1}-\mathbf{k}_{2})/2$ and $%
\mathbf{k}_{\mathrm{ph}}=(\mathbf{k}_{2}+\mathbf{k}_{3})/2$ are the average
momenta in the particle-particle and the particle-hole channels, $\mathbf{q}%
_{\mathrm{ph1}}=\mathbf{k}_{1}-\mathbf{k}_{3}$ and the momentum transfer in
the latter two channels, $\mathbf{q}_{\mathrm{pp}}=\mathbf{k}_{1}+\mathbf{k}%
_{2}$ and $\mathbf{q}_{\mathrm{ph}}=\mathbf{k}_{3}-\mathbf{k}_{2}.$ Using
decomposition (\ref{Dec}), we obtain the equation for the contribution of
the particle-partice channel%
\begin{align}
& \dot{V}_{i_{1..4}}^{\mathrm{pp}}(\mathbf{k,k}^{\prime },\mathbf{q})=
\label{Vpp} \\
& \sum\limits_{i_{1..4},\mathbf{k}^{\prime \prime }}\widetilde{\Pi }%
_{\Lambda ,\mathrm{pp}}^{i_{1}^{\prime }i_{2}^{\prime };i_{3}^{\prime
}i_{4}^{\prime }}(\mathbf{k},\mathbf{q})\left[ V_{i_{1}i_{2}i_{1}^{\prime
}i_{2}^{\prime }}^{\mathrm{pp}}(\mathbf{k,k}^{\prime \prime },\mathbf{q}%
)\right.  \notag \\
& +V_{i_{1}i_{2}i_{1}^{\prime }i_{2}^{\prime }}^{\mathrm{ph}}\left( \frac{%
\mathbf{k-k}^{\prime \prime }+\mathbf{q}}{2},\frac{\mathbf{k}^{\prime \prime
}-\mathbf{k+q}}{2},\mathbf{k}+\mathbf{k}^{\prime \prime }\right)  \notag \\
& \left. +V_{i_{1}i_{2}i_{1}^{\prime }i_{2}^{\prime }}^{\mathrm{ph1}}\left( 
\frac{\mathbf{k+k}^{\prime \prime }+\mathbf{q}}{2},-\frac{\mathbf{k}+\mathbf{%
k^{\prime \prime }-q}}{2},\mathbf{k-k}^{\prime \prime }\right) \right] 
\notag \\
& \times \left[ V_{i_{3}^{\prime }i_{4}^{\prime }i_{3}i_{4}}^{\mathrm{pp}}(%
\mathbf{k}^{\prime \prime },\mathbf{k}^{\prime },\mathbf{q})\right.  \notag
\\
& +V_{i_{3}^{\prime }i_{4}^{\prime }i_{3}i_{4}}^{\mathrm{ph}}\left( \frac{%
\mathbf{k}^{\prime \prime }-\mathbf{k}^{\prime }+\mathbf{q}}{2},\frac{%
\mathbf{k}^{\prime }-\mathbf{k}^{\prime \prime }+\mathbf{q}}{2}\right) 
\notag \\
& \left. +V_{i_{3}^{\prime }i_{4}^{\prime }i_{3}i_{4}}^{\mathrm{ph1}}\left( 
\frac{\mathbf{k}^{\prime }\mathbf{+k}^{\prime \prime }+\mathbf{q}}{2},-\frac{%
\mathbf{k}^{\prime }+\mathbf{k^{\prime \prime }-q}}{2},\mathbf{k}^{\prime
\prime }-\mathbf{k}^{\prime }\right) \right] ,  \notag
\end{align}%
and the particle-hole channel,%
\begin{align}
& \dot{V}_{i_{1..4}}^{\mathrm{ph}}(\mathbf{k,k}^{\prime },\mathbf{q}%
)=\sum\limits_{i_{1..4}^{\prime },\mathbf{k}^{\prime \prime }}\widetilde{\Pi 
}_{\Lambda ,\mathrm{pp}}^{i_{1}^{\prime }i_{2}^{\prime };i_{3}^{\prime
}i_{4}^{\prime }}(\mathbf{k}^{\prime \prime },\mathbf{q}) \\
& \times \left[ V_{i_{1}i_{4}^{\prime }i_{1}^{\prime }i_{4}}^{\mathrm{pp}%
}\left( \frac{\mathbf{k}-\mathbf{k}^{\prime \prime }+\mathbf{q}}{2},\frac{%
\mathbf{k}^{\prime \prime }-\mathbf{k+q}}{2},\mathbf{k}+\mathbf{k}^{\prime
\prime }\right) \right.  \notag \\
& +V_{i_{1}i_{4}^{\prime }i_{1}^{\prime }i_{4}}^{\mathrm{ph}}(\mathbf{k,k}%
^{\prime \prime },\mathbf{q})  \notag \\
& \left. +V_{i_{1}i_{4}^{\prime }i_{1}^{\prime }i_{4}}^{\mathrm{ph1}}(\frac{%
\mathbf{k+k}^{\prime \prime }+\mathbf{q}}{2},\frac{\mathbf{k}+\mathbf{%
k^{\prime \prime }-q}}{2},\mathbf{k-k}^{\prime \prime })\right]  \notag \\
& \times \left[ V_{i_{3}^{\prime }i_{2}i_{3}i_{2}^{\prime }}^{\mathrm{pp}%
}\left( \frac{\mathbf{k}^{\prime \prime }-\mathbf{k}^{\prime }+\mathbf{q}}{2}%
,\frac{\mathbf{k}^{\prime }-\mathbf{k}^{\prime \prime }+\mathbf{q}}{2}%
\right) \right.  \notag \\
& +V_{i_{3}^{\prime }i_{2}i_{3}i_{2}^{\prime }}^{\mathrm{ph}}(\mathbf{k}%
^{\prime \prime },\mathbf{k}^{\prime },\mathbf{q})  \notag \\
& \left. +V_{i_{3}^{\prime }i_{2}i_{3}i_{2}^{\prime }}^{\mathrm{ph1}}\left( 
\frac{\mathbf{k}^{\prime }\mathbf{+k}^{\prime \prime }+\mathbf{q}}{2},\frac{%
\mathbf{k}^{\prime }+\mathbf{k^{\prime \prime }-q}}{2},\mathbf{k}^{\prime
\prime }\mathbf{-k}^{\prime }\right) \right] .  \notag
\end{align}%
We represent the vertex $V^{\mathrm{ph1}}$ according to the Eq. (\ref{Vph1})
of the main text, and in the most part of calculations, we neglect the
dependence of $V^{\mathrm{pp,ph}}$ and $\widetilde{V}^{\mathrm{ph1}}$ on the
third argument (as discussed in the main text), projecting it to zero. We
have verified that treatment of the full momentum dependence does not change
the results substantially. For the vertex $\widetilde{V}^{\mathrm{ph1}}$ we
obtain the equation 
\begin{align}
& \dot{\widetilde{V}}_{i_{1..4}}^{\mathrm{ph1}}(\mathbf{k},\mathbf{k}%
^{\prime },\mathbf{q})%
\begin{array}{c}
=%
\end{array}%
\sum\limits_{i_{1..4}^{\prime },\mathbf{k}^{\prime \prime }}\widetilde{\Pi }%
_{\Lambda ,\mathrm{ph}}^{i_{2}^{\prime }i_{3}^{\prime };i_{1}^{\prime
}i_{4}^{\prime }}(\mathbf{k}^{\prime \prime },\mathbf{q}) \\
& \left\{ \left[ V_{i_{1}i_{1}^{\prime }i_{3}i_{3}^{\prime }}^{\mathrm{pp}%
}\left( \frac{\mathbf{k-k}^{\prime \prime }+\mathbf{q}}{2},\frac{\mathbf{k-k}%
^{\prime \prime }-\mathbf{q}}{2},\mathbf{k}+\mathbf{k}^{\prime \prime
}\right) \right. \right.  \notag \\
& +V_{i_{1}i_{1}^{\prime }i_{3}i_{3}^{\prime }}^{\mathrm{ph}}\left( \frac{%
\mathbf{k}+\mathbf{k}^{\prime \prime }+\mathbf{q}}{2},\frac{\mathbf{k}+%
\mathbf{k}^{\prime \prime }-\mathbf{q}}{2},\mathbf{k}-\mathbf{k}^{\prime
\prime }\right)  \notag \\
& \left. +\widetilde{V}_{i_{1}i_{1}^{\prime }i_{3}i_{3}^{\prime }}^{\mathrm{%
ph1}}(\mathbf{k},\mathbf{k}^{\prime \prime },\mathbf{q})\right]  \notag \\
& \times \left[ V_{i_{4}^{\prime }i_{2}i_{4}i_{2}^{\prime }}^{\mathrm{pp}%
}\left( \frac{\mathbf{k}^{\prime \prime }-\mathbf{k}^{\prime }+\mathbf{q}}{2}%
,\frac{\mathbf{k}^{\prime }-\mathbf{k}^{\prime \prime }+\mathbf{q}}{2},%
\mathbf{k}^{\prime \prime }+\mathbf{k}^{\prime }\right) \right.  \notag \\
& +V_{i_{4}^{\prime }i_{2}i_{4}i_{2}^{\prime }}^{\mathrm{ph}}\left( \mathbf{k%
}^{\prime \prime },\mathbf{k}^{\prime },\mathbf{q}\right)  \notag \\
& \left. +V_{i_{4}^{\prime }i_{2}i_{4}i_{2}^{\prime }}^{\mathrm{ph1}}(\frac{%
\mathbf{k}^{\prime \prime }+\mathbf{k}^{\prime }+\mathbf{q}}{2},\frac{%
\mathbf{k}^{\prime }+\mathbf{k}^{\prime \prime }-\mathbf{q}}{2},\mathbf{k}%
^{\prime \prime }-\mathbf{k}^{\prime })\right]  \notag \\
& +\left[ V_{i_{1}i_{1}^{\prime }i_{3}^{\prime }i_{3}}^{\mathrm{pp}}\left( 
\frac{\mathbf{k-k}^{\prime \prime }+\mathbf{q}}{2},\frac{\mathbf{k}^{\prime
\prime }-\mathbf{k}+\mathbf{q}}{2},\mathbf{k}+\mathbf{k}^{\prime \prime
}\right) \right.  \notag \\
& +V_{i_{4}^{\prime }i_{2}i_{4}i_{2}^{\prime }}^{\mathrm{ph}}\left( \mathbf{k%
},\mathbf{k}^{\prime \prime },\mathbf{q}\right)  \notag \\
& \left. +V_{i_{1}i_{1}^{\prime }i_{3}^{\prime }i_{3}}^{\mathrm{ph1}}\left( 
\frac{\mathbf{k+k}^{\prime \prime }+\mathbf{q}}{2},\frac{\mathbf{k}^{\prime
\prime }+\mathbf{k}-\mathbf{q}}{2},\mathbf{k-k}^{\prime \prime }\right) %
\right]  \notag \\
& \times \left[ V_{i_{4}^{\prime }i_{2}i_{2}^{\prime }i_{4}}^{\mathrm{pp}%
}\left( \frac{\mathbf{k}^{\prime \prime }-\mathbf{k}^{\prime }+\mathbf{q}}{2}%
,\frac{\mathbf{k}^{\prime \prime }-\mathbf{k}^{\prime }-\mathbf{q}}{2},%
\mathbf{k}^{\prime \prime }+\mathbf{k}^{\prime }\right) \right.  \notag \\
& +V_{i_{4}^{\prime }i_{2}i_{2}^{\prime }i_{4}}^{\mathrm{ph}}\left( \frac{%
\mathbf{k}^{\prime \prime }\mathbf{+k}^{\prime }+\mathbf{q}}{2},\frac{%
\mathbf{k}^{\prime \prime }+\mathbf{k}^{\prime }-\mathbf{q}}{2},\mathbf{k}%
^{\prime \prime }-\mathbf{k}^{\prime }\right)  \notag \\
& \left. +\widetilde{V}_{i_{4}^{\prime }i_{2}i_{2}^{\prime }i_{4}}^{\mathrm{%
ph1}}(\mathbf{k}^{\prime \prime },\mathbf{k}^{\prime },\mathbf{q})\right] 
\notag
\end{align}%
\begin{align*}
& -2\left[ V_{i_{1}i_{1}^{\prime }i_{3}i_{3}^{\prime }}^{\mathrm{pp}}\left( 
\frac{\mathbf{k-k}^{\prime \prime }+\mathbf{q}}{2},\frac{\mathbf{k-k}%
^{\prime \prime }-\mathbf{q}}{2},\mathbf{k}+\mathbf{k}^{\prime \prime
}\right) \right. \\
& +V_{i_{1}i_{1}^{\prime }i_{3}i_{3}^{\prime }}^{\mathrm{ph}}\left( \frac{%
\mathbf{k}+\mathbf{k}^{\prime \prime }+\mathbf{q}}{2},\frac{\mathbf{k}+%
\mathbf{k}^{\prime \prime }-\mathbf{q}}{2},\mathbf{k}-\mathbf{k}^{\prime
\prime }\right) \\
& \left. +\widetilde{V}_{i_{1}i_{1}^{\prime }i_{3}i_{3}^{\prime }}^{\mathrm{%
ph1}}(\mathbf{k},\mathbf{k}^{\prime \prime },\mathbf{q})\right] \\
& \times \left[ V_{i_{4}^{\prime }i_{2}i_{2}^{\prime }i_{4}}^{\mathrm{pp}%
}\left( \frac{\mathbf{k}^{\prime \prime }-\mathbf{k}^{\prime }+\mathbf{q}}{2}%
,\frac{\mathbf{k}^{\prime \prime }-\mathbf{k}^{\prime }-\mathbf{q}}{2},%
\mathbf{k}^{\prime \prime }+\mathbf{k}^{\prime }\right) \right. \\
& +V_{i_{4}^{\prime }i_{2}i_{2}^{\prime }i_{4}}^{\mathrm{ph}}\left( \frac{%
\mathbf{k}^{\prime \prime }\mathbf{+k}^{\prime }+\mathbf{q}}{2},\frac{%
\mathbf{k}^{\prime \prime }+\mathbf{k}^{\prime }-\mathbf{q}}{2},\mathbf{k}%
^{\prime \prime }-\mathbf{k}^{\prime }\right) \\
& \left. \left. +\widetilde{V}_{i_{4}^{\prime }i_{2}i_{2}^{\prime }i_{4}}^{%
\mathrm{ph1}}(\mathbf{k}^{\prime \prime },\mathbf{k}^{\prime },\mathbf{q})%
\right] \right\} .
\end{align*}%
Finally, the vertex $\gamma $ and the inverse propagator $\Pi $ are
determined from 
\begin{align}
\dot{g}& _{\mathbf{k},\mathbf{q}}^{i_{2}i_{4}}%
\begin{array}{c}
=%
\end{array}%
\sum\limits_{i_{1..4}^{\prime },\mathbf{k}^{\prime \prime }}\widetilde{\Pi }%
_{\Lambda ,\mathrm{ph}}^{i_{2}^{\prime }i_{3}^{\prime };i_{1}^{\prime
}i_{4}^{\prime }}(\mathbf{k}^{\prime \prime },\mathbf{q}) \\
& \times \left\{ V_{i_{1}i_{1}^{\prime }i_{3}^{\prime }i_{3}}^{\mathrm{pp}%
}\left( \frac{\mathbf{k-k}^{\prime \prime }+\mathbf{q}}{2},\frac{\mathbf{k}%
^{\prime \prime }-\mathbf{k}+\mathbf{q}}{2},\mathbf{k}+\mathbf{k}^{\prime
\prime }\right) \right.  \notag
\end{align}%
\begin{eqnarray*}
&+&V_{i_{4}^{\prime }i_{2}i_{4}i_{2}^{\prime }}^{\mathrm{ph}}\left( \mathbf{k%
},\mathbf{k}^{\prime \prime },\mathbf{q}\right) \\
&+&V_{i_{1}i_{1}^{\prime }i_{3}^{\prime }i_{3}}^{\mathrm{ph1}}\left( \frac{%
\mathbf{k+k}^{\prime \prime }+\mathbf{q}}{2},\frac{\mathbf{k}^{\prime \prime
}+\mathbf{k}-\mathbf{q}}{2},\mathbf{k-k}^{\prime \prime }\right) \\
&-&2\left[ V_{i_{1}i_{1}^{\prime }i_{3}i_{3}^{\prime }}^{\mathrm{pp}}\left( 
\frac{\mathbf{k-k}^{\prime \prime }+\mathbf{q}}{2},\frac{\mathbf{k-k}%
^{\prime \prime }-\mathbf{q}}{2},\mathbf{k}+\mathbf{k}^{\prime \prime
}\right) \right. \\
&+&V_{i_{1}i_{1}^{\prime }i_{3}i_{3}^{\prime }}^{\mathrm{ph}}\left( \frac{%
\mathbf{k}+\mathbf{k}^{\prime \prime }+\mathbf{q}}{2},\frac{\mathbf{k}+%
\mathbf{k}^{\prime \prime }-\mathbf{q}}{2},\mathbf{k}-\mathbf{k}^{\prime
\prime }\right) \\
&+&\left. \left. \widetilde{V}_{i_{1}i_{1}^{\prime }i_{3}i_{3}^{\prime }}^{%
\mathrm{ph1}}(\mathbf{k},\mathbf{k}^{\prime \prime },\mathbf{q})\right]
\right\} g_{\mathbf{k}^{\prime \prime },\mathbf{q}}^{i_{4}^{\prime
},i_{2}^{\prime }}
\end{eqnarray*}%
and%
\begin{equation}
\dot{\Pi}(\mathbf{q})=-2\sum\limits_{i_{1..4}^{\prime },\mathbf{k}^{\prime
\prime }}\widetilde{\Pi }_{\Lambda ,\mathrm{ph}}^{i_{2}^{\prime
}i_{3}^{\prime };i_{1}^{\prime }i_{4}^{\prime }}(\mathbf{k}^{\prime \prime },%
\mathbf{q})g_{\mathbf{k}^{\prime \prime },\mathbf{q}}^{i_{1}^{\prime
},i_{3}^{\prime }}g_{\mathbf{k}^{\prime \prime },\mathbf{q}}^{i_{4}^{\prime
}i_{2}^{\prime }}.  \label{dPi}
\end{equation}%
The equations (\ref{Vpp})-(\ref{dPi}) have to be solved numerically.

Yet, the functions $V_{i_{1..4},\Lambda }^{\mathrm{pp},\mathrm{ph}}$ and $%
\widetilde{V}_{i_{1..4},\Lambda }^{\mathrm{ph1}}$ of two continuum $2$%
-component variables is hard to treat accurately numerically. To treat
accurately these dependences, we follow the idea of Ref. \cite{Salmhofer1},
expanding these functions in some harmonics. Since the dependence on the
absolute value of first two arguments is expected to be singular (as follows
from the singular behavior of the gap function $\Delta (\mathbf{k)\sim |k|}%
^{-\gamma },$ $\gamma \sim 1/2$ on the ordered side of the transition, cf.
Refs. \cite{Murthy,Gusynin1}), we expand in Fourier harmonics with respect
to the angle of each of the two momenta: 
\begin{align}
V_{i_{1..4},\Lambda }^{m}(\mathbf{k,k}^{\prime })& =\sum\limits_{m=0}^{n_{F}}%
\left[ F_{i_{1..4}}^{mm^{\prime },\Lambda }(k,k^{\prime })\cos (m\varphi _{%
\mathbf{k}})\cos (m^{\prime }\varphi _{\mathbf{k}^{\prime }})\right.  \notag
\label{Vharm} \\
& +G_{i_{1..4}}^{mm^{\prime },\Lambda }(k,k^{\prime })\cos (m\varphi _{%
\mathbf{k}})\sin (m^{\prime }\varphi _{\mathbf{k}^{\prime }})  \notag \\
& +H_{i_{1..4}}^{mm^{\prime },\Lambda }(k,k^{\prime })\sin (m\varphi _{%
\mathbf{k}})\cos (m^{\prime }\varphi _{\mathbf{k}^{\prime }})  \notag \\
& \left. +J_{i_{1..4}}^{mm^{\prime },\Lambda }(k,k^{\prime })\sin (m\varphi
_{\mathbf{k}})\sin (m^{\prime }\varphi _{\mathbf{k}^{\prime }})\right]
\end{align}%
and discretizing the absolute values of the momenta, with further performing
(bi-)linear interpolation between the discretization points. In this way, we
achieve sufficiently fine discretization of each vertex, (typically we take $%
n_{F}=2$ Fourier components, corresponding to $2n_{F}+1=5$ Fourier
harmonics, and $n=20\div 25$ radial points logarithmically distributed in
the range $[\Lambda _{\min }/10,\Lambda _{\text{uv}}];$ we take $\Lambda _{%
\mathrm{uv}}=2$ and $\Lambda _{\mathrm{\min }}=e^{-10}$). The results of the
solution of Eqs. (\ref{Vpp})-(\ref{Vharm}) are discussed in the main text.

\end{document}